\documentclass[%
reprint,
amsmath,amssymb,
aps,
]{revtex4-2}

\usepackage{graphicx}
\usepackage{dcolumn}
\usepackage{bm}

\graphicspath{{./F/}}
\usepackage{physics}
\def\den{\rho}

\def\ra{\rightarrow}
\def\eps{\epsilon}
\def\mc{\multicolumn}
\def\D{\Delta}
\def\half{\frac{1}{2}}
\def\TF{^{\rm TF}}
\def\sss{\scriptscriptstyle\rm}
\def\F{_{\sss F}}
\def\floor#1{{\lfloor}#1{\rfloor}}
\def\AEAtwo{^{\rm AEA2}}

\def\asech{\text{asech}}
\def\ben{\begin{equation}}
\def\een{\end{equation}}

\def\1D{^{\rm 1D}}

\usepackage{color}
\usepackage[colorlinks=true,linkcolor=blue,citecolor=blue]{hyperref}

\usepackage{mathrsfs} 

\begin{document}

\preprint{APS/123-QED}

\title{Orbital-free functional with sub-milliHartree errors for slabs}

\author{Pavel Okun}
\affiliation{Department of Chemistry, University of California, Irvine, CA 92617,  USA}
\author{Antonio C. Cancio}
\affiliation{Department of Physics and Astronomy, Ball State University, Muncie, IN 47304, USA}
\author{Kieron Burke}
\affiliation{Departments of Physics and Astronomy and of Chemistry, University of California, Irvine}

\date{\today}

\begin{abstract}
Using principles of asymptotic analysis, we derive the exact leading corrections to the Thomas-Fermi kinetic energy approximation for Kohn-Sham electrons for slabs.  This asymptotic expansion approximation includes crucial quantum oscillations missed by standard semilocal density functionals.  Because these account for the derivative discontinuity, chemical accuracy is achieved at fourth-order.  The implications for both orbital-free electronic structure and exchange-correlation approximations are discussed.
\end{abstract}

\maketitle

\section{Introduction}
Almost all modern density functional theory (DFT) calculations use the Kohn-Sham (KS) scheme \cite{KS65}, where only the exchange-correlation (XC) energy is approximated as a density functional \cite{DG90}.  Accuracy comes at the cost of solving self-consistently the KS equations for the orbitals.  However, the KS kinetic energy, $T_s$, is also a density functional \cite{DG90,YP94}. If it could be approximated with sufficient accuracy, without incurring substantial additional computational cost, one would bypass the KS equations \cite{LC05}, speeding up every DFT calculation on the planet.  The dream of orbital-free DFT lives on \cite{MLTP23}.

Thomas-Fermi (TF) theory \cite{T27,F27}, the original DFT, is orbital-free, but is too crude for modern calculations.  First, $T_s$ is typically far larger than the XC energy, so a far smaller fractional error is required.  Second, via the Euler equation that follows from the variational principle \cite{L83}, the functional derivative of $T_s$ determines the density, which is typically very poor \cite{KSB13}.

Over the decades, many attempts have been made to construct orbital-free density functional approximations \cite{DG90,LC05,FSCS22,WW13,WRDC18}, often aimed at a limited set of circumstances \cite{LKT18}.  These include the original gradient expansion from slowly-varying densities \cite{K57,SP99,DG90,Y86}, generalizations of that expansion \cite{LKT18,CFLS11}, functionals designed for weakly interacting subsystems such as water molecules \cite{LKW08}, two-point functionals for use in materials calculations \cite{WGC99,HC10}, and functionals for surfaces \cite{SMP21}.  

But Lieb and Simon (LS) showed long ago \cite{LS73,LS77} that, in a very specific semiclassical limit, TF theory becomes relatively exact.  One approach to the LS limit is to take $\hbar\to 0$, keeping $\mu$, the chemical potential, fixed. Expanding about this limit, functionals of the potential have been developed for the total energy of one-dimensional (1D) problems \cite{BB20,B20,B20b,OB21b}.  The dominant term is given by a TF calculation, while higher order terms include not only the gradient expansion, but also oscillating corrections to it, and depend on both the Maslov \cite{MF01} indices and whether the system is finite or extended.  By resumming this asymptotic series for a linear half-well, the total energy of 10 non-interacting fermions was found to 33 digits \cite{BB20}.

\begin{figure}[htb!]
\includegraphics[width=0.95\columnwidth]{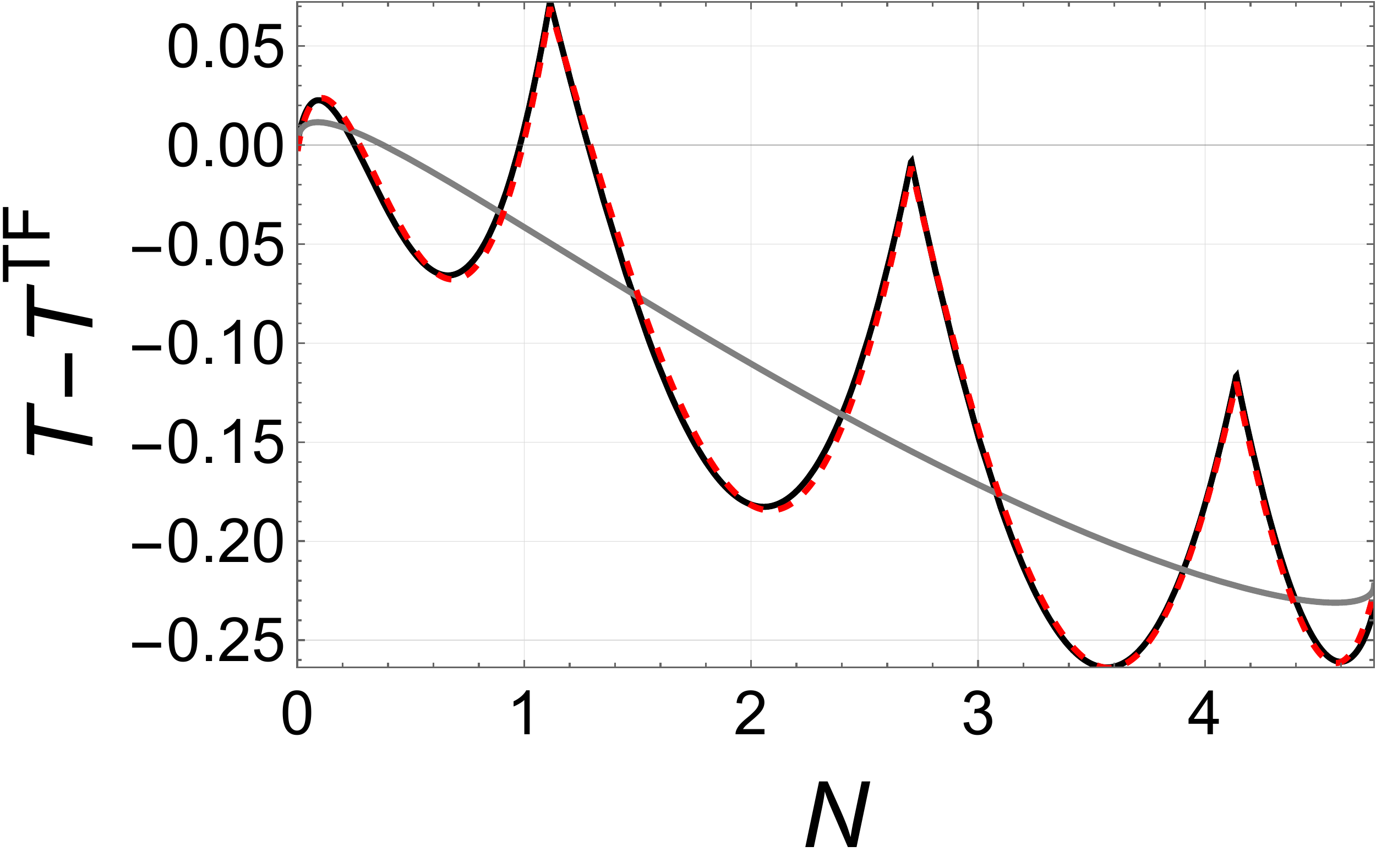}
\caption{Slab kinetic energy per unit area vs electron number per unit area, with the exact (black), gradient expansion (gray), and asymptotic approximation (red, see text).  The TF kinetic energy has been subtracted from all curves.  Cusps, i.e., derivative discontinuities, occur when a new band starts to be filled.}
\label{fig:TNL}
\end{figure}

Here, we demonstrate the capabilities of a potential functional to yield chemical accuracy [errors below 1 milliHartree (mH)] for a three-dimensional slab geometry, with a potential that varies in only one direction.  We find excellent results when the leading correction is added to TF theory, and chemical accuracy when the next two orders are included, providing a systematic, parameter-free approach to orbital-free calculations.  Figure \ref{fig:TNL} illustrates the deviation of the kinetic energy from its TF value, for a Pöschl-Teller (PT) slab with well depth $10$.  The standard density functional, the gradient expansion approximation (GEA) yields the smooth gray curve, which averages over the density of states.  Our potential functional (red curve) includes the quantum oscillations characteristic of systems with discrete states, with errors almost too small to be visible here.  These results are characteristic of any system to which our functional applies, and it is unlikely any existing orbital-free {\em density} functional can compete.  Our calculation (a) shows how the asymptotic expansion corrects standard semi-local functionals to account for the derivative discontinuity \cite{SCY08,KK20}, (b) connects the rich field of semiclassical analysis of  eigenvalue problems, including densities of states and the importance of Maslov indices \cite{BB03,G90}, to the construction of approximate functionals for non-trivial 3D problems, (c) explicitly connects the failures of semilocal functionals for stretched bonds (symmetry breaking) to the divergence of the asymptotic expansion \cite{CSY08}, and (d) provides insight for creating functionals of chemical accuracy for systems with more general potentials \cite{CCKB18}.

\section{Background}
For any quantum problem, one can consider functionals of either the potential or the density (thanks to Hohenberg-Kohn \cite{HK64}).  The KS scheme requires {\em density} functionals for the XC energy, but orbital-free approximations can use either variable for $T_s$.  The derivation of the gradient expansion of DFT begins with a potential functional approximation (PFA) for a slowly-varying gas \cite{K57}, and converts it, order-by-order, into a density functional approximation (DFA) \cite{DG90}.  The lowest order is Thomas-Fermi theory, which is local in either the density or the potential.  For a slowly-varying gas, the procedure is extended to include second-order gradients in either variable.  The oscillating contributions captured by the PFAs derived here depend on integrals over the entire system, i.e., they have no known analog as density functionals.
 
Before discussing slab PFAs, our main topic in this paper, we will briefly discuss PFAs in 1D.  This will give us the opportunity to introduce our notation and semiclassical jargon which may be unfamiliar to nonspecialists.  Moreover for non-interacting particles, one dimension is special, because the semiclassical expansion of the eigenvalues can be performed to arbitrary order (the celebrated WKB expansion \cite{BO99}).  The relation to DFT was first noted in 1956, when (non-interacting) TF theory was derived by summing WKB eigenvalues \cite{MP56}.  Lastly we will use 1D quantities to construct our slab functionals.

Just as in Ref. \cite{BB20}, we define a quantum action as a function that yields the eigenvalues $\eps_j$ via \cite{BO99}
\begin{equation}
\label{act}
s(\eps_j) = j + \nu, \qquad j=0, 1, \cdots,
\end{equation}
where $\nu$ is the Maslov index.  In this paper we only consider smooth single potential wells so $\nu = 1/2$.  Expanding $s$ in powers of $\hbar$ yields a series expansion for $s$ in even gradients of the potential \cite{BO99}:
\begin{equation}
\label{sExpand}
s=s^{(0)}+\D s^{(2)}+\D s^{(4)} + \cdots.
\end{equation}
We use atomic units ($\hbar = m_e = 1$), so energies are in Hartrees and distances are in Bohr radii.  In these units the lowest order action (also called the classical action) at energy $\mu$ is 
\begin{equation}
\label{s0}
s^{(0)}(\mu) = \int dx\, \frac{p\F(x)}{\pi},
\end{equation}
where $p\F(x) = \sqrt{2[\mu - v(x)]}$ is the classical momentum of a particle with energy $\mu$.  All spatial integrals throughout this paper are understood to run only between the classical turning points satisfying $v(\pm x\F) = \mu$, as we consider only symmetric single well potentials $v(x)$.  Inserting Eq. (\ref{s0}) into Eq. (\ref{act}) yields the familiar WKB eigenvalues found in introductory textbooks on quantum mechanics \cite{G05}.

The higher-order corrections are well-defined, although less well-known.   A useful way of writing them was given in Ref. \cite{KLR67}.    Define
\begin{align}
\label{IJ}
\begin{split}
I(\mu) &= \int dx\, \frac{v''(x) p\F(x)}{8\pi},\\
J(\mu) &= \int dx \frac{7 v''(x)^2 - 5 v^{(4)}(x) p\F(x)^2}{\pi\, p\F(x)},\\
\end{split}
\end{align}
where $v^{(j)}(x) = d^j v/dx^j$.  $I(\mu)$ has a second-order gradient and $J(\mu)$ has fourth order gradients.  Then the next corrections can be written as
\begin{equation}
\label{sHO}
\D s^{(2)}(\mu) = -\frac{I''(\mu)}{3}, \qquad
\D s^{(4)}(\mu) = \frac{J'''(\mu)}{5760},
\end{equation}
where the derivatives on $I$ and $J$ are with respect to $\mu$.  This particular way of writing the higher order action terms avoids dealing with divergences at the turning points (compared to the forms in Ref. \cite{BO99}).

The 1D number staircase $n(\mu)$ gives the number of eigenstates below $\mu$:
\begin{equation}
\label{nTheta}
n(\mu) = \sum_{j = 0}^{\infty} \Theta(\mu - \eps_j),
\end{equation}
where $\Theta(x)$ is the Heaviside theta function.  In terms of the quantum action \cite{B20}
\begin{equation}
\label{n}
n(\mu) = s(\mu) - \expval{s(\mu)},
\end{equation}
where $\expval{y} = y - \floor{y + 1/2}$ is periodic, oscillating between $\pm 1/2$, and $\floor{y}$ is the integer part of $y$.

The exact 1D total energy can be written as a {\em potential} functional using the 1D number staircase:
\begin{equation}
\label{E1}
E(\mu) = \mu\, n(\mu) - \int_{0}^{\mu} d\eps\, n(\eps).
\end{equation}
Using the asymptotic expansion of $s$ in Eq. (\ref{sExpand}) we can construct the asymptotic expansion of the 1D number staircase.  Later we will use this expansion of $n(\mu)$ to construct order by order approximations for our slab potential functionals.  We define the asymptotic expansion approximation (AEA) to order $M$ (AEAM) as
\begin{equation}
\label{nAEAM}
n^{(M)}(\mu) = s^{(M)}(\mu) - \expval{s^{(M)}(\mu)}, \qquad {\rm AEAM}
\end{equation}
i.e., it includes terms up to $M$-th order in both the smooth and oscillating pieces.  Because the oscillating contribution is bounded in the LS limit, where $j \to \infty$ in Eq. (\ref{act}), it is typically of lower order than the smooth contribution.  Thus we define an alternative to AEA, denoted AEA', as
\begin{equation}
\label{nAEAMp}
n^{(M)}(\mu) = s^{(M)}(\mu) - \expval{s^{(M-2)}(\mu)}, \qquad {\rm AEAM'}
\end{equation}
which is easier to calculate and is correct to the same order.  However, any derivative with respect to $\mu$ may not be.

\begin{figure}[htb!]
\includegraphics[width=0.95\columnwidth]{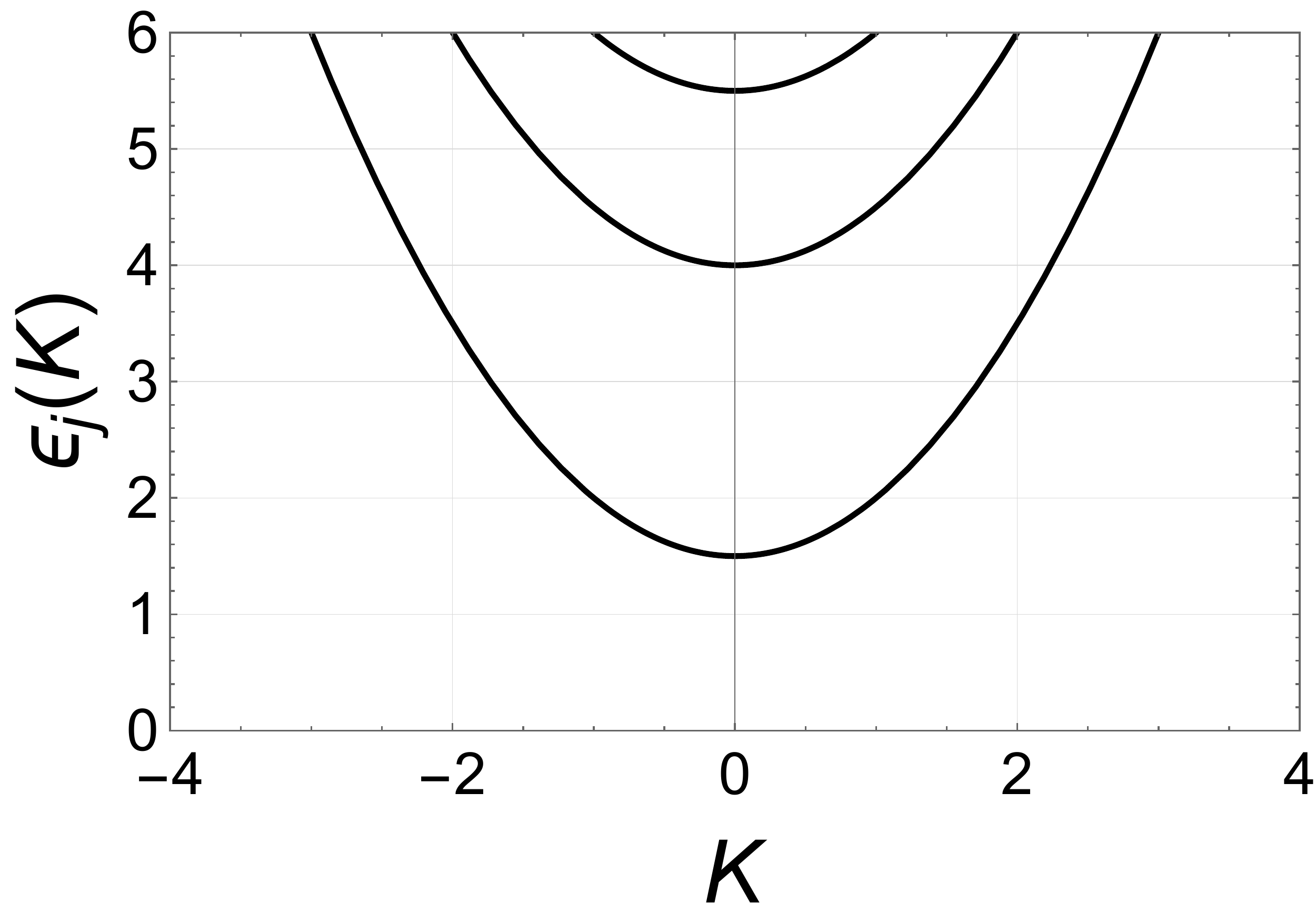}
\caption{The eigenvalues of a slab with a Pöschl-Teller well of depth 6 as a function of the parallel wave vector $K$.}
\label{fig:epsK}
\end{figure}

We have shown in Eq. (\ref{E1}) how the 1D number staircase determines the total energy.  It also determines the kinetic and potential energy and the density.  If you have $n(\mu)$ you have all the information you need about a system.  This is not surprising since Eq. (\ref{nTheta}) shows that knowing $n$ is equivalent to knowing all the eigenvalues.  The structure of $n(\mu)$, in particular its oscillatory term, is very interesting.  If we ignore the oscillating term entirely and approximate $n(\mu)$ with the lowest-order action, $n(\mu) \approx s^{(0)}(\mu)$, we recover 1D TF theory.  If we use a higher-order action, $n(\mu) \approx s^{(M)}(\mu)$, we recover the 1D GEA to $M$-th order \cite{SP99}.  Thus TF theory corresponds to the special case $M = 0$, without the oscillating contribution.  Here we derive the GEA as a potential functional, but it is always possible to invert it into the traditional density functional GEA.  However, the failure of the GEA, when applied to finite systems, is well known \cite{MB68}.  In this case (single well 1D potentials),  we can easily see why the GEA fails: it neglects the oscillatory term $\expval{s(\mu)}$.  By including this term we derive PFAs which (unlike the GEA) yield the correct asymptotic expressions.  The resulting PFAs are much more accurate, and they capture the derivative discontinuity.  But converting PFAs into DFAs is much more difficult when these oscillatory terms are included, and it may not even be possible.  The lesson is clear: the GEA fails to produce correct asymptotics because it ignores this oscillatory piece.  Modern generalized gradient approximations (GGAs) that change the GEA coefficient \cite{CFLS11} cannot compensate for this missing oscillatory correction in general, so either their accuracy or their range of applicability is limited.  Restoration of the oscillations is crucial on a path to a systematic, asymptotically correct, expansion of functional approximations.

\section{Theory}
Now we apply this technology to a three-dimensional non-interacting problem, a slab with a potential profile $v(x)$, but uniform and infinite in the other two directions.  This geometry allows comparison with any existing orbital-free kinetic energy functional \cite{CFLS11}.  By choosing a $v(x)$ that is the sum of two potential wells (a dimer), separated by a distance $R$, we mimic certain aspects of more realistic calculations, such as binding energies as a function of bond lengths in diatomics \cite{PSB95}.  We will only examine symmetric single well potentials, $v(x) = v(-x)$, whose derivatives are finite up to at least the fourth derivative.  For simplicity we fix $v(0) = 0$.

The eigenstates are two-dimensional free electron bands, with energies $\eps_j(K) = \eps_j+K^2/2$, where $\eps_j$ is the $j$-th eigenvalue in the 1D well, and $K$ is the parallel wavevector.  Figure \ref{fig:epsK} shows the simple shape of the PT slab bands, which are free-electron like in the two directions perpendicular to $v(x)$ and begin at each of the eigenvalues of the 1D well.  The system is a band metal.  Summing over parallel directions in the continuum limit, with double occupation, yields the number of electrons per unit area from the 1D number staircase:
\begin{equation} 
\label{NE}
N(\mu) = \int_{0}^{\mu} \frac{d\eps}{\pi} n(\eps),
\end{equation}
i.e., the density of states per unit area, $dN/d\mu$, is simply proportional to the 1D number staircase.  The slab energy (per unit area) is then
\begin{equation}
\label{E}
E(\mu) = \mu\, N(\mu) - \int_{0}^{\mu} d\eps\, N(\eps).
\end{equation}
From now on all energy components are per unit area.  Since $N(\mu)$ is completely determined by $n(\mu)$ the slab $E(\mu)$ is also completely determined by $n(\mu)$.  Figure \ref{fig:TNL} shows its kinetic contribution.  The density of states, d$N/d\mu$, consists of constants (2D uniform gas), with steps up at each $\eps_j$.  We populate our slabs with non-interacting KS electrons.  Inversion of $N(\mu)$ then yields $E(N)$, either for the exact functional or any approximation.  The density is never calculated if not needed for other purposes.

Inserting $s^{(0)}$ and dropping the oscillatory term in Eq. (\ref{n}), and then plugging this $n(\mu)$ into Eqs. (\ref{NE}) and (\ref{E}) yields the familiar (non-interacting) TF theory to lowest order:
\begin{align}
\begin{split}
\label{TF}
N\TF(\mu) &= \int \frac {dx}{3\pi^2}\, p\F(x)^3,\\
E\TF(\mu) &=  \int \frac {dx}{3\pi^2}\, \left[  \frac{3}{10} p\F(x)^2 + v(x) \right] p\F(x)^3.\\
\end{split}
\end{align} 
Both of these are potential functionals, and the first term in the energy is the kinetic contribution.  As we typically solve problems for a given $N$, one inverts to find $\mu\TF(N)$ and then $E\TF(N)=E\TF(\mu\TF(N))$.

To extract the density from such a PFA, we take the functional derivative of the energy functional with respect to the potential while keeping the particle number $N$ fixed: 
\begin{equation}
\den\TF(x) = \frac{\delta E\TF}{\delta v(x)} \bigg|_N = \frac{p\F(x)^3}{3\pi^2}.
\end{equation}
We could also use this to eliminate $p\F(x)$ from the above expressions, thereby converting to the usual density functional approximations.  The same equations result from starting with the kinetic energy as a density functional, and finding the density by minimizing the total energy for fixed $N$.

A principal achievement of this work is the derivation of the leading post-TF corrections.   There is no correction to first order, but even orders appear to have both smooth and oscillatory terms, while third order is purely oscillatory.  A full careful derivation, which we will give elsewhere, involves subtle interchanges of orders of limits to avoid apparent divergences, but a non-rigorous derivation is straightforward.   The smooth contributions in Eq. (\ref{n}) from $s(\mu)$ can be easily integrated, as they are given by explicit derivatives in Eq (\ref{sHO}).  The subtleties occur in the oscillating contributions.   These can be treated by replacing the integration variable $\eps$ in Eq. (\ref{NE}) with the action itself.  The change of variables allows the integrand to be evaluated by elementary means (as in Eq. (16) of Ref. \cite{B20}), but only if $s'(\mu)$ is known as a function of $s(\mu)$.  This smooth function can be found, order-by-order, by expanding for large $\mu$.   The resulting expression for the slab staircase is
\begin{equation}
\label{N2}
\D N^{(2)} = - \frac{I'(\mu)}{3\pi} + \frac{q}{2\tau},
\end{equation}
where
\begin{equation}
\tau = \pi s^{(0) \prime}(\mu) = \int \frac{dx}{p\F(x)},
\end{equation}
is the classical time to cross the well at energy $\mu$, and
\begin{equation}
\label{q}
q = \frac{1}{12} - \expval{s}^2,
\end{equation}
is a periodic function of the action, originating from an integral proportional to the linear term in Eq. (16) of Ref. \cite{B20}.  The smooth contribution can be isolated by taking $\tau \ra \infty$.  These smooth terms yield the standard gradient expansion when converted to density functionals.

The total energy can then be deduced via the same procedure we used to find $N(\mu)$, but applied to Eq. (\ref{E}), yielding
\begin{equation}
\label{E2}
\D E^{(2)} = \mu\, \D N^{(2)}  +\frac{I}{3\pi}.
\end{equation}
Given an expression for the total energy, one can extract the potential contribution by taking a derivative with respect to the magnitude of $v(x)$.  We do this order-by-order in the 1D expressions, and find the slab contribution via
\begin{equation}
V = \int_{0}^{\mu} \frac{d\eps}{\pi} V\1D(\eps),
\end{equation}
which can be subtracted from the total energy, to yield
\begin{equation}
\label{T2}
\D T^{(2)} = -\frac{I}{6\pi}+ \frac{\pi s^{(0)} q}{4\tau^2}.
\end{equation}
These expressions apply to {\em any} smooth symmetric $v(x)$ with one minimum and two turning points.  As we illustrate below, these are the exact asymptotic corrections to TF theory for such problems.  The red curve in Fig. \ref{fig:TNL} is the AEA2 kinetic energy.

The process is repeated to find further higher-order terms, in which the lengthy algebra and careful avoidance of apparent divergences, will be reported elsewhere.  Here we simply report the results:
\begin{align}
\label{N3E3T3}
\begin{split}
\D N^{(3)} &= \frac{\pi \tau'}{6 \tau^3} h, \qquad
\D E^{(3)} = \mu \D N^{(3)} - \frac{\pi h}{6 \tau^2},\\
\D T^{(3)} &= \frac{\pi^2 s^{(0)} \tau'}{4 \tau^4} h,\\
\end{split}
\end{align}
where all primes are derivatives with respect to $\mu$ and
\begin{equation}
h = \expval{s}\left(q + \frac{1}{6} \right) . 
\end{equation}
The fourth order corrections are
\begin{align}
\label{N4E4T4}
\begin{split}
\D N^{(4)} &=  \frac{J''}{5760 \pi} + \frac{\pi I'''}{6 \tau^2} q -
\frac{\pi^2 w}{2 \tau^4} \left[ 3 \frac{\tau'^2}{\tau} - \tau'' \right],\\
\D E^{(4)} &= \mu\, \D N^{(4)} - \frac{J'}{5760 \pi} + \frac{3 \pi^2 \tau'}{2 \tau^4} w,\\
\D T^{(4)} &= - \frac{J'}{11520 \pi} + \frac{\pi q}{12 \tau^2} \left[ I'' + \frac{ 2 \pi s^{(0)} I'''}{\tau} \right] +\\
& \frac{\pi^2 w}{4 \tau^4} \left[3 \tau' + 4\pi s^{(0)} \frac{\tau''}{\tau} -15 \pi s^{(0)} \left( \frac{\tau'}{\tau} \right)^2 \right],\\
\end{split}
\end{align}
with $w = (7 - 240 g)/2880$ and 
\begin{equation}
g = \expval{s}^2 \left( q + \frac{5}{12} \right).
\end{equation}
We note that our separation of terms into smooth and oscillatory is based on their origin in the derivation, rather than any process averaging over energy levels.

\section{Pöschl-Teller slabs}
Our next step is to apply these formulas to a specific problem.  Our prototypical well is the Pöschl–Teller well, with $v(x) = D\, \tanh^2 x$.  Such potentials are used in semiconductor physics \cite{RMII00,HP17}, and yield analytic formulas.  Here $s^{(0)} = \sqrt{2D}c$, $\D s^{(2)} = 1/(8\sqrt{2D})$, and $\D s^{(4)} = - 1/(256 \sqrt{2D^3})$, where $c = 1 - \sqrt{1 - \mu/D}$.  The TF PT particle number and energies are
\begin{align}
\begin{split}
N\TF &=  \frac{\sqrt{2D^3} c^2}{\pi} \left(1 - \frac{2}{3}c \right),\\
E\TF &= \frac{\sqrt{2D^5} c^3}{\pi} \left(\frac{4}{3} - \frac{3 c}{2} + \frac{2 c^2}{5} \right),\\
T\TF &= \frac{3}{2} (\mu N\TF-E\TF).\\
\end{split}
\end{align}
The 2nd order contributions from Eqs. (\ref{N2}), (\ref{E2}), and (\ref{T2}) are
\begin{align}
\begin{split}
\D N^{(2)} &= \frac{\sqrt{2D}}{48\pi} [c\, (4 - 3c) - 24 \expval{s}^2 (1 - c)],\\
\D T^{(2)} &= - \frac{\sqrt{2D^3}}{192 \pi} [c^2 (4 - 3c) + 96\expval{s}^2 (1 - c)^2] c,\\
\D E^{(2)} &= \mu \left[\D N^{(2)} + \frac{\sqrt{2D}}{96\pi} (4 - 6c + 3c^2) \right].\\
\end{split}
\end{align}
The third order corrections from Eq. (\ref{N3E3T3}) are
\begin{align}
\begin{split}
\D N^{(3)} &= \frac{h}{6 \pi}, \\ \D E^{(3)} &= \mu \D N^{(3)} - \frac{D h}{3 \pi} (1 - c)^2,\\
\D T^{(3)} &= \frac{D h}{2 \pi} c (1 - c).\\
\end{split}
\end{align}
The fourth order corrections from Eq. (\ref{N4E4T4}) are
\begin{align}
\begin{split}
\D N^{(4)} &= \frac{2 - 6 c + 3 c^2}{768 \pi \sqrt{2D}},\\
\D E^{(4)} &= \mu \D N^{(4)} + \frac{\sqrt{2D}}{2\pi} \beta,\\ 
\D T^{(4)} &= \frac{\sqrt{2D}}{4\pi} \left[ \beta - \frac{(1 - c)^2 q + 24 c w}{8} \right],\\
\end{split}
\end{align}
where
\begin{equation}
\beta = \frac{24 - 40 c + 80 c^2 - 60 c^3 + 15 c^4}{7680} + 3 w (1 - c).
\end{equation}

\begin{figure}[htb!]
\includegraphics[width=0.95\columnwidth]{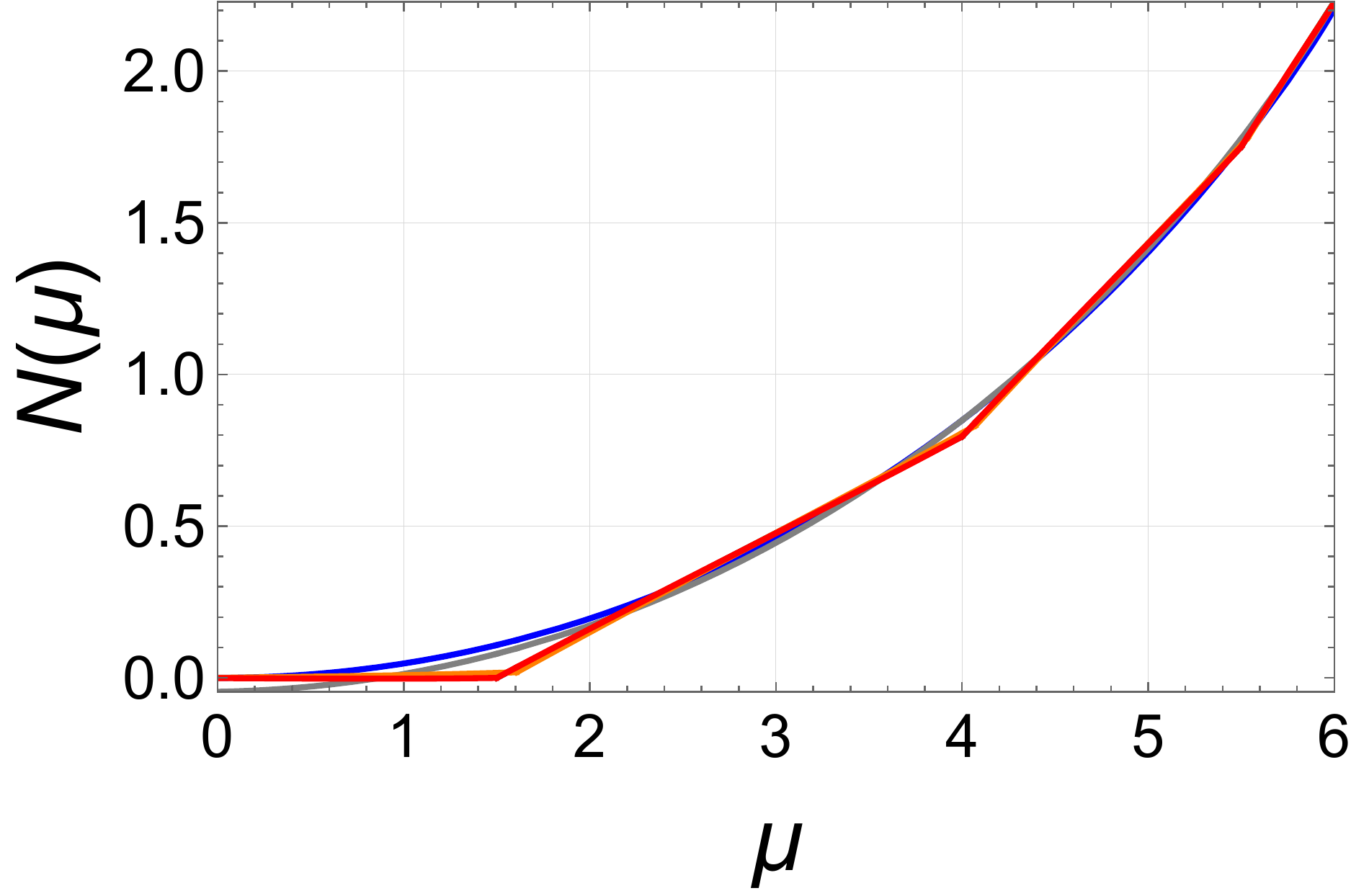}
\includegraphics[width=0.95\columnwidth]{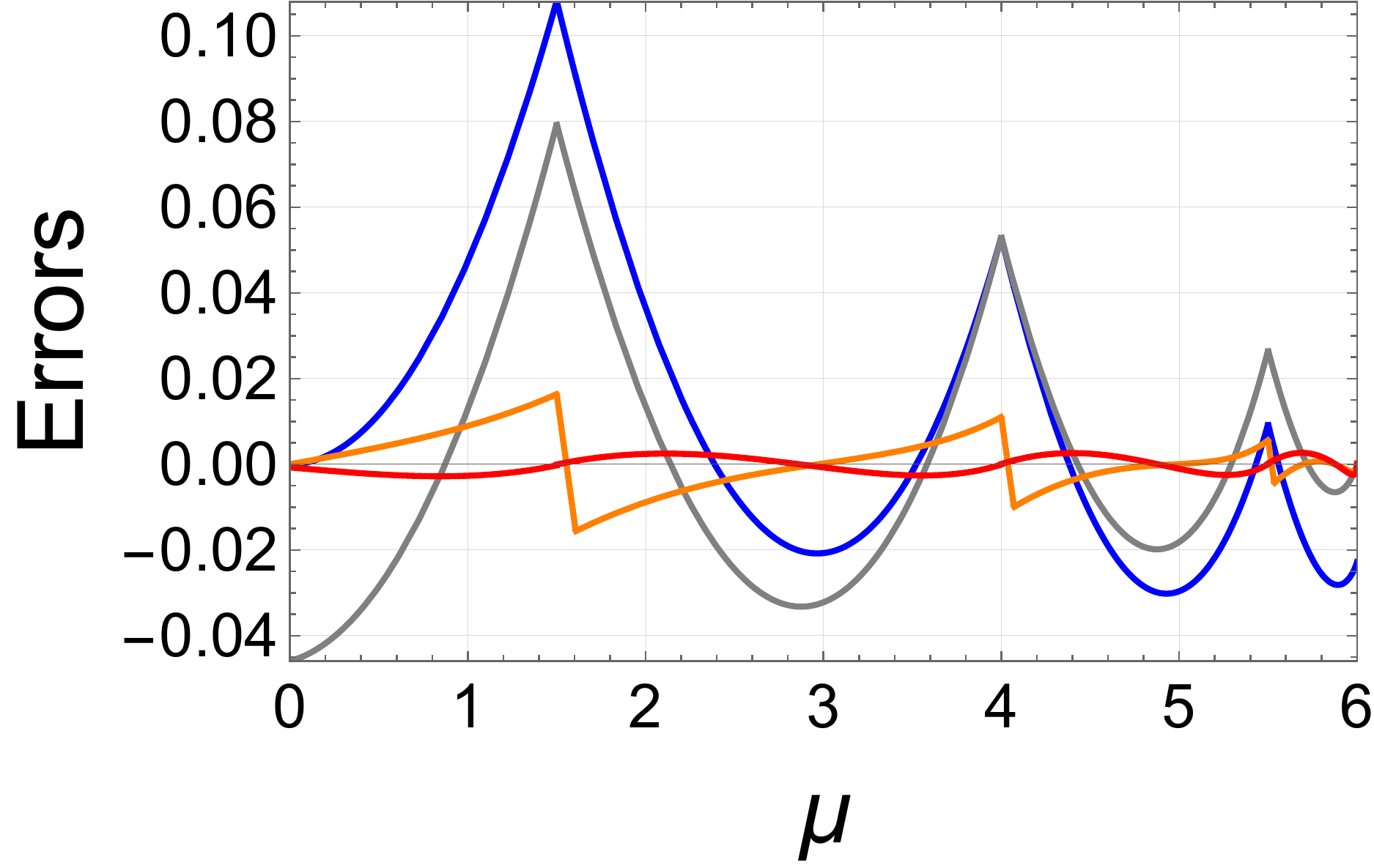}
\caption{The PT slab number staircase with $D = 6$ (black), and its TF (blue), GEA2 (gray), AEA2' (orange), and AEA2 (red) approximations.  The lower panel shows the errors in the upper panel.}
\label{fig:N}
\end{figure}

\label{sec:bands}
Figure \ref{fig:N} shows the number staircase (integrated density of states) for a given PT slab.   The function is rather smooth, making it difficult to see differences between approximations.  However, there are kinks in the exact curve whenever a new band begins to be occupied.   Both the TF and GEA2 curves have no such kinks.
Figure \ref{fig:N} also plots the errors (defined as approximate minus exact) in the number staircase of the various approximations.  The TF and GEA2 error curves have kinks because they are smooth, while the exact curve is not.  The orange curve is AEA2', which only accounts for the leading behavior of the oscillating term, while the red curve is AEA2, which includes the next contribution to the oscillating term.  Their difference becomes negligible for sufficiently large $\mu$ (both are asymptotically correct), but AEA2 clearly has smaller errors for small $\mu$.

\begin{figure}[htb!]
\includegraphics[width=0.95\columnwidth]{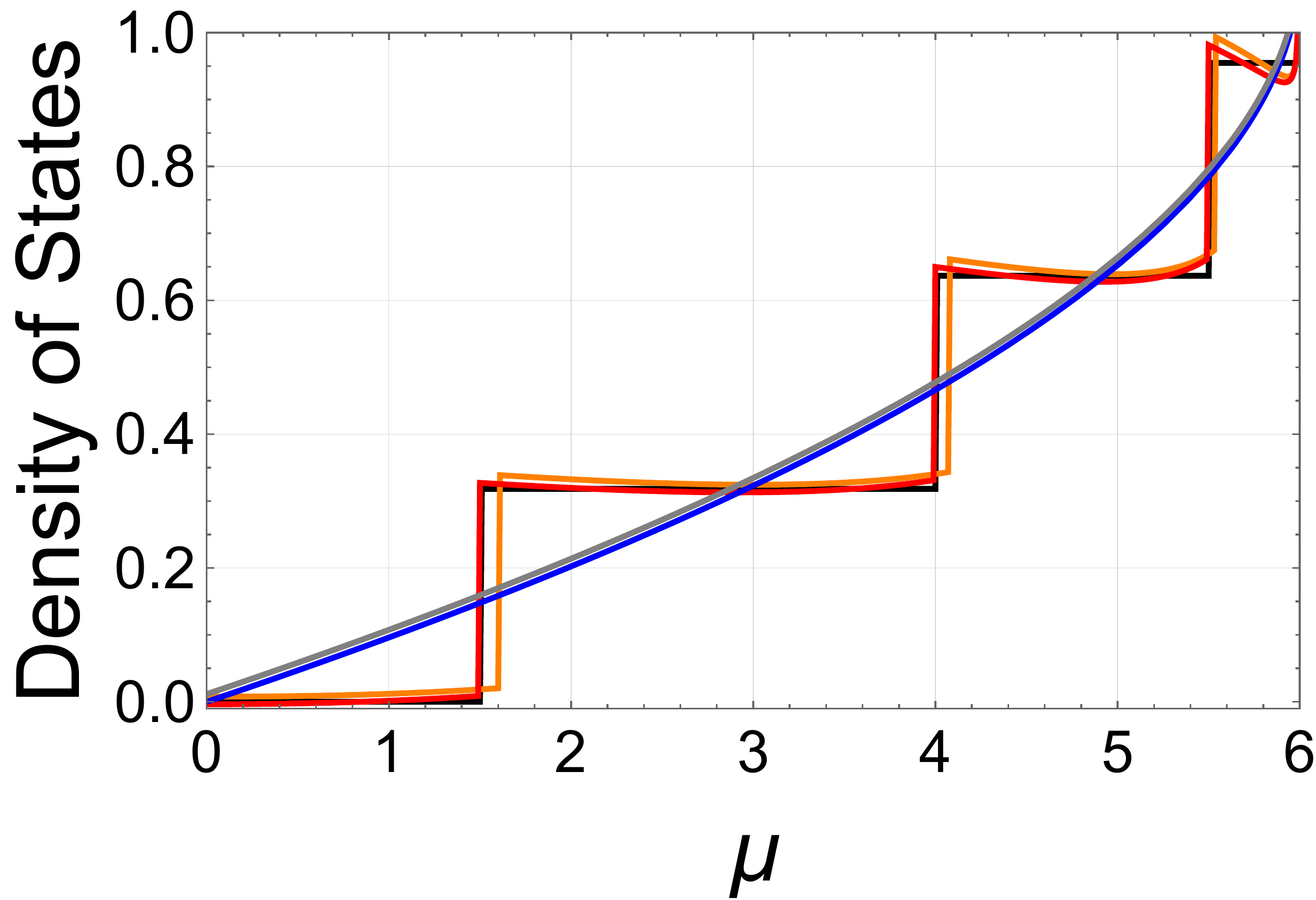}
\includegraphics[width=0.95\columnwidth]{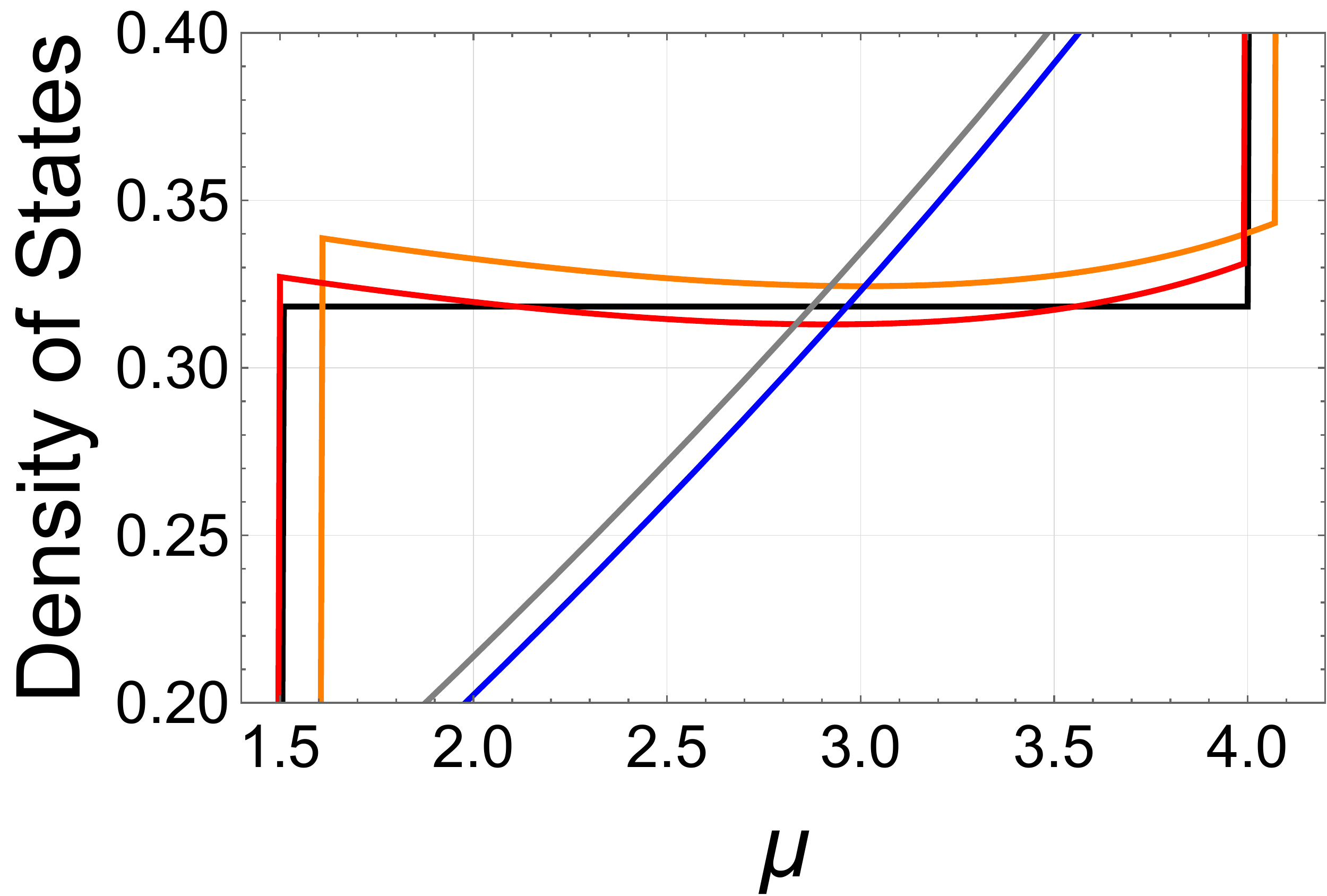}
\caption{The, $D = 6$, PT slab density of states $dN/d\mu$ (black), and its TF (blue), GEA2 (gray), AEA2' (orange), and AEA2 (red) approximations.}
\label{fig:DOS}
\end{figure}
Figure \ref{fig:DOS} shows the density of states of a particular PT slab.  This is just the derivative of the number staircase given in Eq. (\ref{NE}) and shown in Fig. \ref{fig:N}.  Both TF and GEA2 yield smooth approximations to it, and miss the discrete steps (the origin of the infamous DFT derivative discontinuity \cite{SB20}).  Unlike how it is treated in many semiclassical works \cite{BB03}, the smooth curve is not synonymous with the TF contribution, as GEA2 makes a small but finite correction.  The asymptotic expansion approximation contains approximate steps, with approximations to the plateau inbetween.  The exact density of states jumps discontinuously when $\mu = \eps_j$, where the $\eps_j$ are the exact 1D eigenvalues.  Using the definition of the saw-tooth function $\expval{x} = x - \floor{x + 1/2}$, we can see that the AEA2' approximation jumps whenever
\begin{equation}
s^{(0)}(\mu) = j + \half, \qquad j = 0,1,2,\cdots.
\end{equation}
This is just the lowest order WKB quantization rule for a single 1D well \cite{G05}.  This means that AEA2' jumps when $\mu = \eps^{(0)}_j$, the $j$th WKB eigenvalue.  Similar analysis shows that AEA2 jumps when
\begin{equation}
s^{(2)}(\mu) = j + \half, \qquad j = 0,1,2,\cdots,
\end{equation}
which is just the second order WKB quantization rule from Eq. (\ref{act}).  Thus AEA2 jumps when $\mu = \eps^{(2)}_j$, the second order WKB eigenvalue.  AEA2 is much more accurate, as the lower panel of Fig. \ref{fig:DOS} shows, because $\eps^{(2)}_j$ is a better approximation to $\eps_j$ than $\eps^{(0)}_j$.  The inaccuracies in both second order AEAs vanish as $\mu$ becomes large.  Neither curve is quite flat, but AEA2 is flatter than AEA2'.

\label{sec:dens}
\begin{figure}[htb!]
\includegraphics[width=0.95\columnwidth]{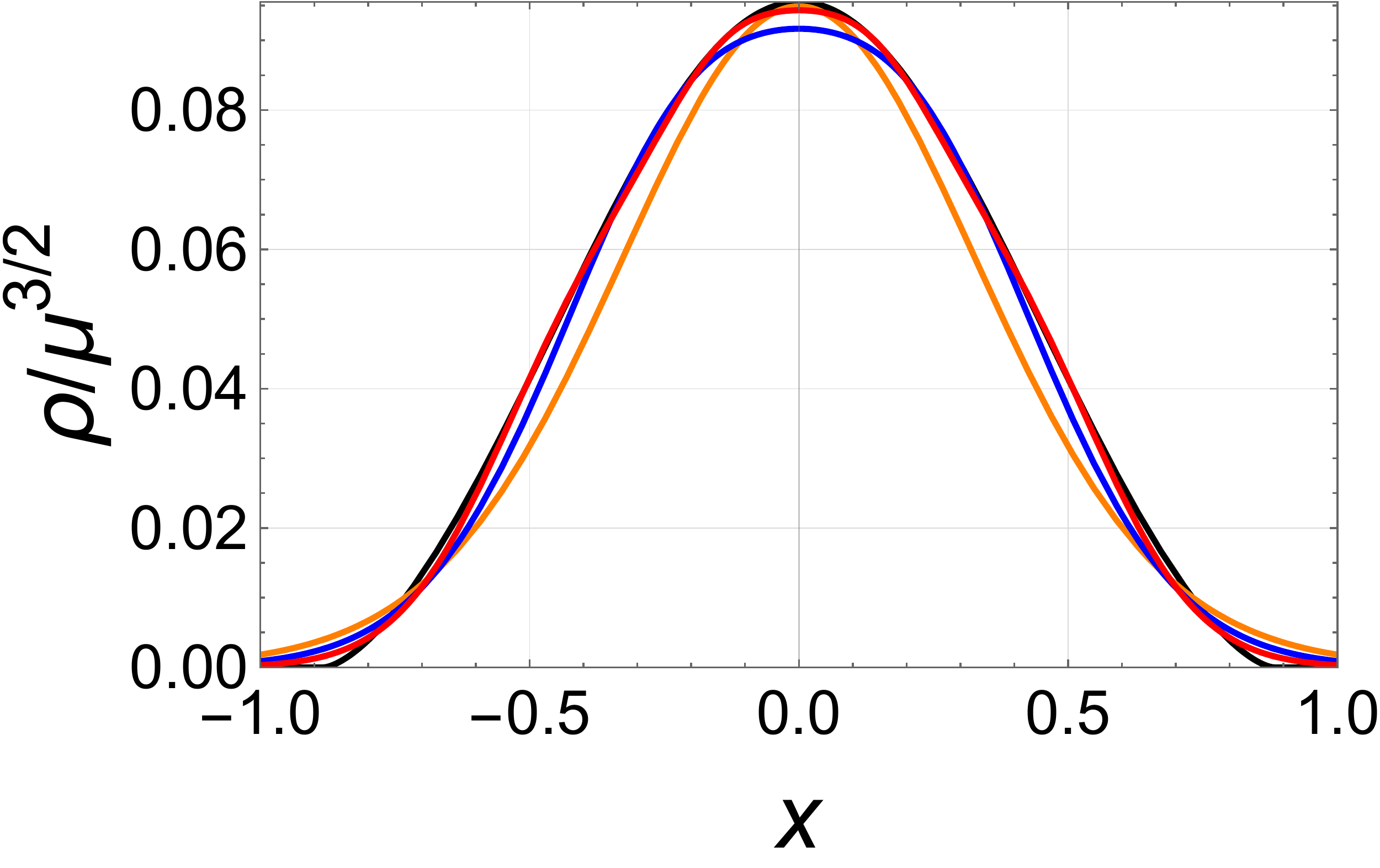}
\includegraphics[width=0.95\columnwidth]{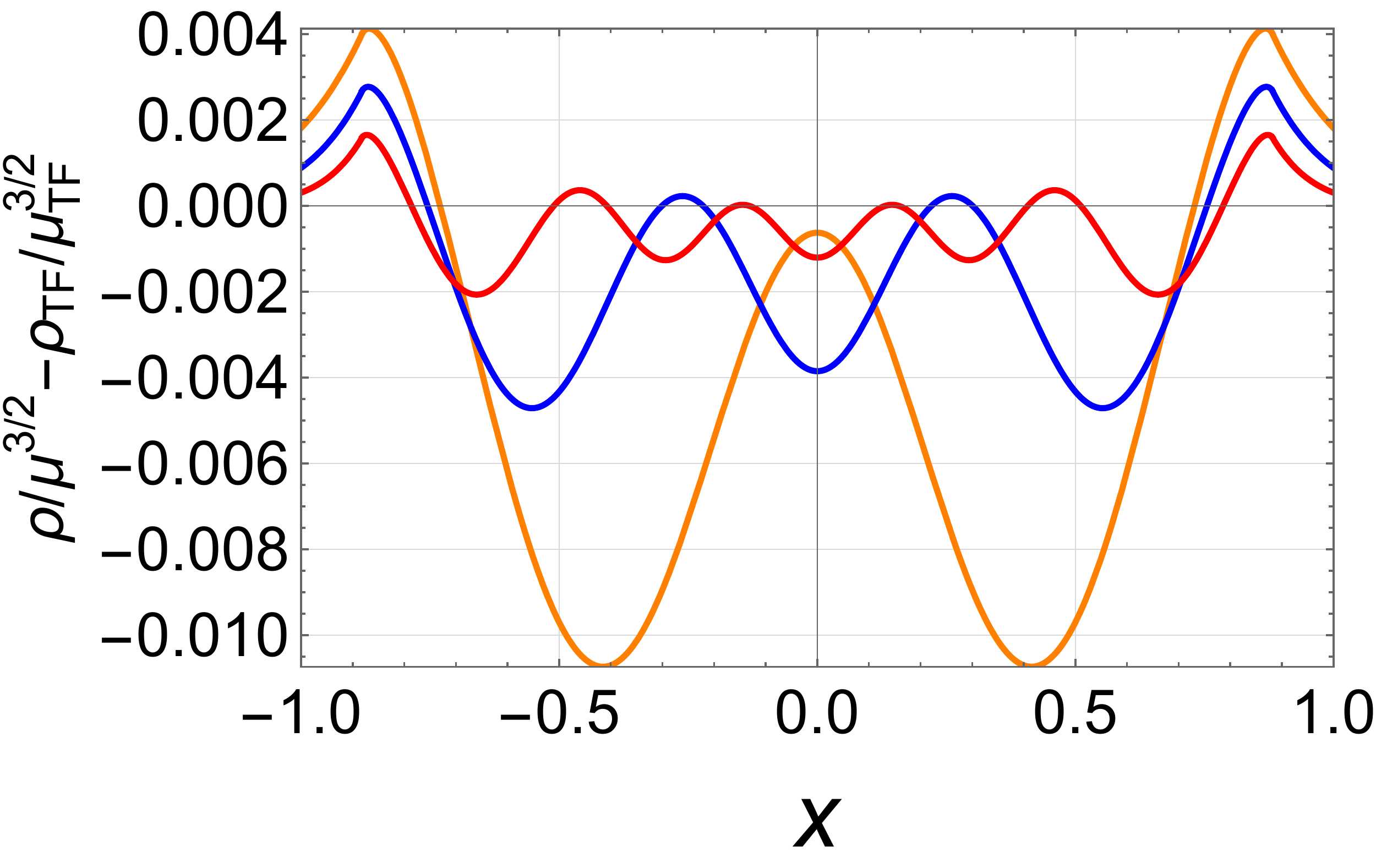}
\caption{Upper panel: Densities for $M = 1$ (orange), 2 (blue), 4 (red), and their TF limiting value (black).  The areas under the curves are TF (0.0879), M = 1 (0.0809), M = 2 (0.0854), M = 4 (0.0871).  Lower panel: The deviation of the exact scaled densities from the scaled TF density.  The exact chemical potential is $\mu$ and $\mu_{\rm{TF}}$ is its TF approximation.}
\label{fig:PTScaledDen}
\end{figure}

\begin{table*}
$\begin{array}{|c|r|r|r|c|c|c|c|r|c|c|c|c|}
\hline
\mc{4}{|c|}{} & \mc{9}{c|}{\text{Errors (mH)}}\\
\hline
\mc{4}{|c|}{} & \mc{5}{c|}{\text{Potential Functionals}} & \mc{4}{c|}{\text{Density Functionals}}\\
\hline
M & \mc{1}{c|}{D} & \mc{1}{c|}{N} & \mc{1}{c|}{T/N} & \text{TF} & \text{GEA2} & \text{AEA2'} & \text{AEA2} & \mc{1}{c|}{\text{AEA4}} & \text{TF} & \text{GEA2} & \text{MGE2} & \text{GEA4} \\
\hline
1  & 12.685 &   1.293 &   3.059 & -87 & -126 & -29 & -2.74 & 0.04097 & -156 & -41 & -8 & -2 \\
2  & 36.000 &   6.525 &   8.728 & -85 & -125 & -14 & -0.92 & 0.00487 & -159 & -35 &  1 & -6 \\
3  & 70.971 &  18.318 &  17.233 & -85 & -125 &  -9 & -0.46 & 0.00124 & -162 & -31 &  7 & -7 \\
4  &117.599 &  39.293 &  28.573 & -84 & -125 &  -7 & -0.28 & 0.00045 & -164 & -28 & 11 & -6 \\
5  &175.883 &  72.075 &  42.748 & -84 & -125 &  -6 & -0.18 & 0.00020 & -165 & -26 & 14 & -6 \\
6  &245.824 & 119.288 &  59.757 & -84 & -125 &  -5 & -0.13 & 0.00010 & -166 & -25 & 16 & -6 \\
7  &327.422 & 183.555 &  79.602 & -84 & -125 &  -4 & -0.10 & 0.00006 & -167 & -24 & 18 & -6 \\
8  &420.677 & 267.500 & 102.281 & -84 & -125 &  -3 & -0.08 & 0.00003 & -168 & -23 & 19 & -6 \\
9  &525.589 & 373.746 & 127.795 & -84 & -125 &  -3 & -0.06 & 0.00002 & -168 & -22 & 21 & -5 \\
10 &642.157 & 504.918 & 156.145 & -84 & -125 &  -3 & -0.05 & 0.00001 & -169 & -21 & 22 & -5 \\
\hline
\end{array}$
\caption{Kinetic energy per particle for Pöschl-Teller slabs with $D= 2\eps_M$ and $\mu = D/2$.  Here AEA$p$ is the $p$-th order asymptotic expansion approximation, GEA$p$ is its analog without oscillating terms, while MGE2 is a generalized gradient approximation (GGA) \cite{CFLS11}.  The rightmost columns are density functionals evaluated on the exact density.  Table \ref{tab:PTWDMuDConE} in the appendix shows the analogous results for the total energy.}
\label{tab:PTWDMuDCon}
\end{table*}

\section{Numerical results}
\begin{figure}[htb!]
\includegraphics[width=0.95\columnwidth]{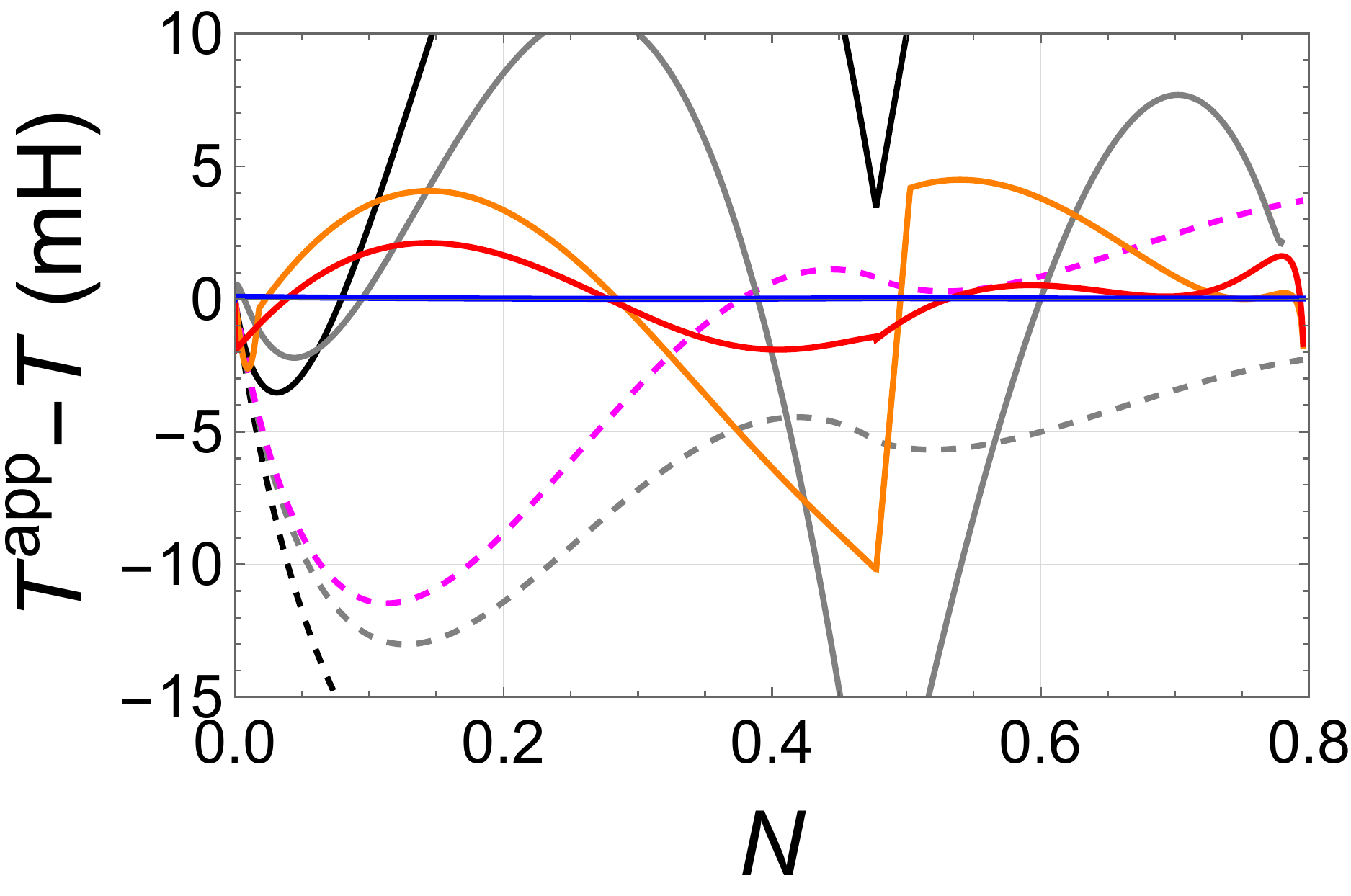}
\caption{Kinetic energy errors for a shallow PT slab ($D = 3$). Legend: TF (black), GEA2 (gray), MGE2 (magenta), AEA2' (orange), AEA2 (red), AEA4 (blue).  Dashed lines denote density functionals acting on the exact density.}
\label{fig:PTSTN}
\end{figure}

In Table \ref{tab:PTWDMuDCon} we report results for a PT slab \cite{N78}.  We approach the LS limit by deepening the well while keeping $\mu/D = 1/2$ fixed.  Values of $D$ are chosen where a new band begins to fill at $\mu = \eps_j$.  Fig. \ref{fig:PTSTN} shows that our approximations are poorest when $\mu$ is close to a 1D eigenvalue, but still  beat the competition.  For each PFA, $N(\mu)$ is found to the given order and inverted numerically to yield $E[\mu(N)] = E(N)$.  Here $T/N$ is the exact kinetic energy per particle in Hartrees, but errors are in milliHartrees.  In this paper errors are defined as approximate minus exact.  The TF PFA error starts at about 3\%, and shrinks to less than 0.1\%, consistent with the LS theorem \cite{LS73,LS77}.  TF is particularly good for our slab, due to the smoothness of the density in the $x$-direction and uniformity in the other two (see Fig. \ref{fig:PTScaledDen}).  The (2nd-order) gradient expansion (GEA2) PFA worsens the result, but AEA2 has much smaller errors, especially for larger $D$.  In fact, errors are less than a mH for all but the shallowest well.

\begin{table*}
$\begin{array}{|c|r|r|c|c|c|c|c|c|c|}
\hline
\mc{3}{|c|}{} & \mc{7}{c|}{\text{Errors (mH)}}\\
\hline
\mc{3}{|c|}{} & \mc{3}{c|}{e_{RE}} & \mc{4}{c|}{\text{$\mu$}}\\
\hline
M & \mc{1}{c|}{e_{RE}} & \mc{1}{c|}{\mu}  & \text{TF} & \text{AEA2} & \text{AEA4} & \text{TF} & \text{AEA2'} & \text{AEA2} & \text{AEA4} \\
\hline
1  &   5.557 &   6.342 &  -63 &   -3 & 0.185 & -242 & -41 & 0.010 & -0.0011911 \\
2  &  17.607 &  18.000 & -172 & -131 & 0.296 & -239 & -21 & 0.013 & -0.0001006 \\
3  &  35.224 &  35.486 & -204 & -166 & 0.159 & -238 & -14 & 0.009 & -0.0000218 \\
4  &  58.603 &  58.799 & -217 & -180 & 0.095 & -237 & -11 & 0.006 & -0.0000072 \\
5  &  87.784 &  87.941 & -224 & -187 & 0.062 & -237 &  -9 & 0.004 & -0.0000030 \\
6  & 122.781 & 122.912 & -227 & -191 & 0.044 & -237 &  -7 & 0.003 & -0.0000015 \\
7  & 163.599 & 163.711 & -230 & -194 & 0.033 & -237 &  -6 & 0.003 & -0.0000008 \\
8  & 210.240 & 210.339 & -231 & -196 & 0.025 & -237 &  -5 & 0.002 & -0.0000005 \\
9  & 262.707 & 262.794 & -232 & -197 & 0.020 & -237 &  -5 & 0.002 & -0.0000003 \\
10 & 321.000 & 321.079 & -233 & -198 & 0.016 & -237 &  -4 & 0.001 & -0.0000002 \\
\hline
\end{array}$
\caption{The removal energies, $e_{RE}$, and the chemical potentials for the slabs of Table \ref{tab:PTWDMuDCon}.  The GEA2, AEA2', and AEA2 total energy PFA's, as functions of $N$, are identical and thus yield the same $e_{RE}$ values.}
\label{tab:IPAndMuTrunc}
\end{table*}

The right-hand set of columns are DFAs evaluated on the exact density.  Unusually, TF does better self-consistently than on the exact density \cite{VSKS19,SSVB22}, but GEA2 is improved.  The MGE2 \cite{CFLS11} (gradient expansion with coefficient modified to recover asymptotics of atoms) is better for small wells, but comparable to GEA2 in the asymptotic limit.  GEA4 is better, but not nearly as good as AEA2.  AEA4 is the best approximation and always yields sub-milliHartree errors, which are microHartree for $M > 3$.

This table only tests one value of $\mu/D$, but similar trends hold for any value.  In Fig. \ref{fig:PTSTN}, we plot errors for a shallow PT slab, $D=3$, as a function of $N$, far from the LS limit.  Our 2nd order approximations are typically better than GEA2 and even MGE2 in this range, and all are beaten by fourth order.  Both this figure and Fig. \ref{fig:TNL} show that the AEA includes cusps in the kinetic energy as a function of $N$, the infamous derivative discontinuity that is the source of many errors in modern DFT calculations \cite{SCY08,KK20}.  Cusps occur when $\mu = \eps_j$ both exactly and in approximations.  The resemblance with Fig. 9 of Ref. \cite{BCGP16} (non-interacting electrons in a Coulomb potential) is striking.

Figure \ref{fig:PTScaledDen} shows exact densities, from Table \ref{tab:PTWDMuDCon}, and their TF approximation.  We scaled the densities so that the TF density is the same for all values of $M$.  As $M$ increases we approach the semiclassical limit and these densities weakly approach their TF counterpart.  Because the chemical potential (relative to well-depth) is held fixed but the particle number is not (unlike in Fig. 2 of Ref. \cite{CLEB10}), the normalization changes, but approaches that of TF in the LS limit.  The lower panel simply shows the differences from the TF curve, making the weak approach to TF evident.  Here, weak means that the integral over any well-behaved function times the exact density approaches its TF counterpart \cite{L76}.

\section{Energy differences}
While it is important to demonstrate the validity of our expressions on total energies, essentially all useful DFT calculations are of energy differences, such as ionization potentials and binding energies of molecules (infinitesimal differences determine bond lengths and lattice parameters).  In the asymptotic behavior of ionization potentials of neutral atoms \cite{CSPB10}, local exchange in a KS calculation was found to match the exact result (Hartree-Fock), even capturing variations across a row of the (very extended) non-relativistic periodic table.  Moreover, the average over such a row matched that of extended interacting TF theory.

\subsection{Removal Energies}
\begin{figure}[htb!]
\includegraphics[width=0.75\columnwidth]{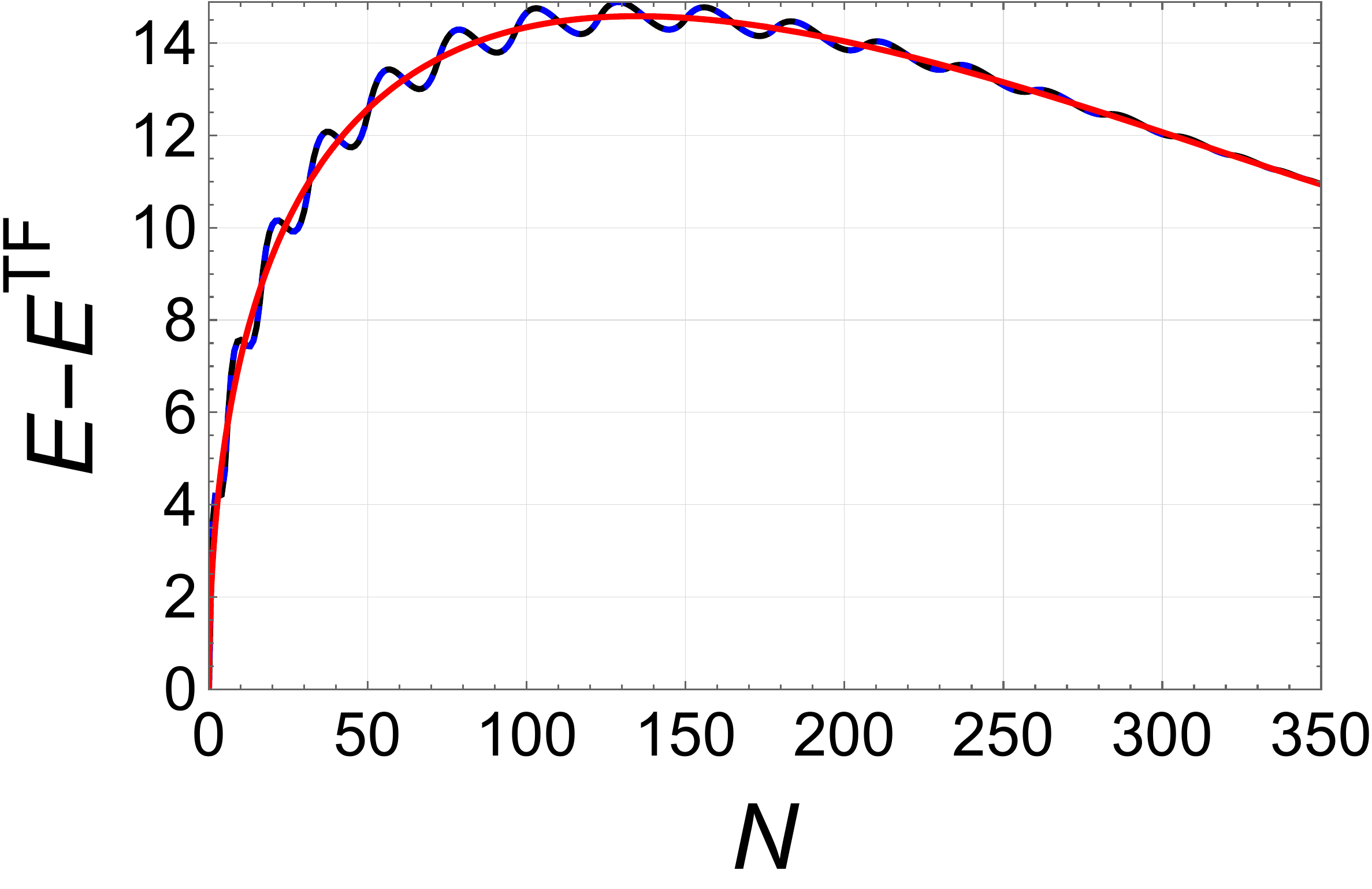}
\includegraphics[width=0.75\columnwidth]{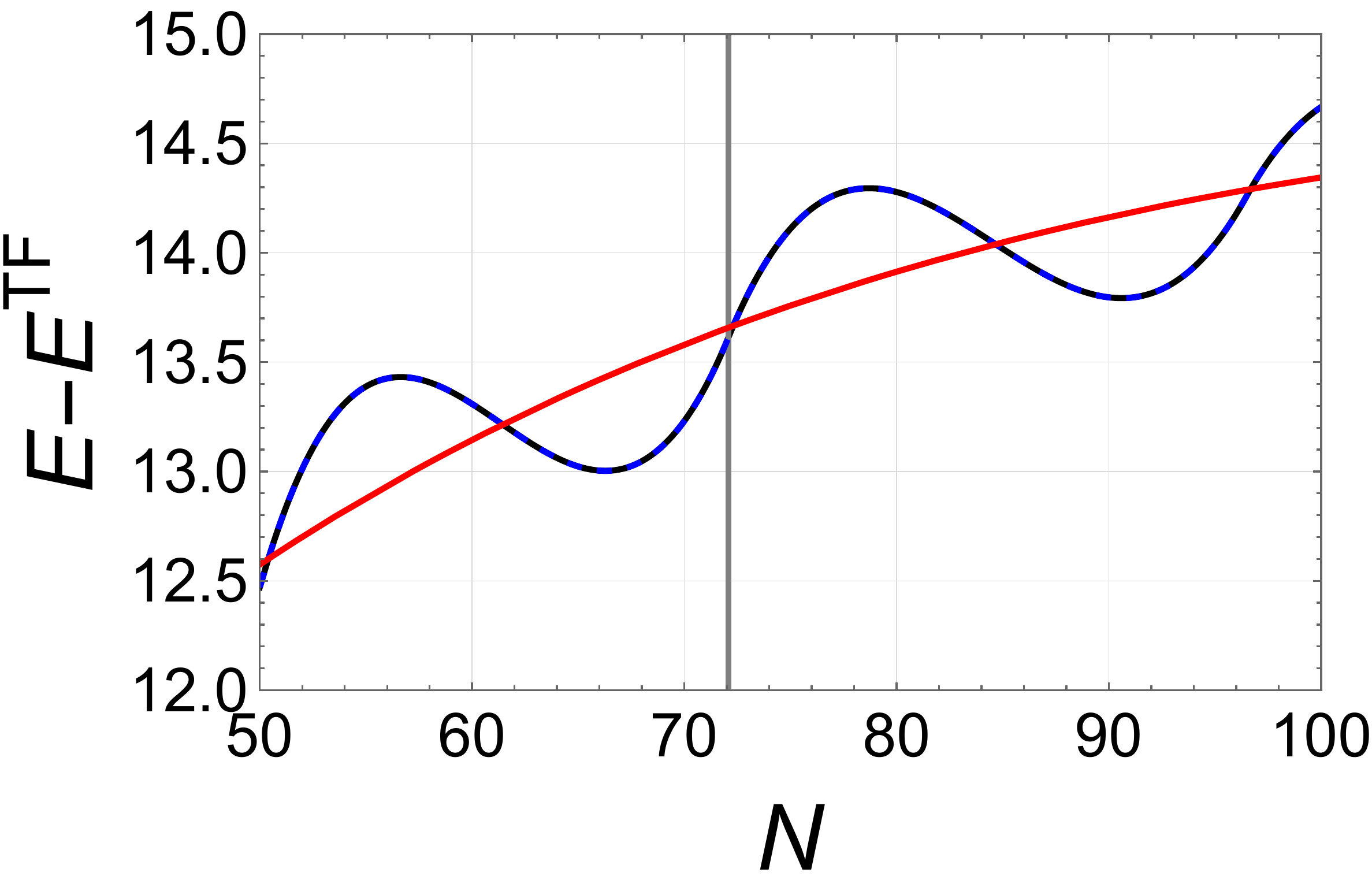}
\includegraphics[width=0.75\columnwidth]{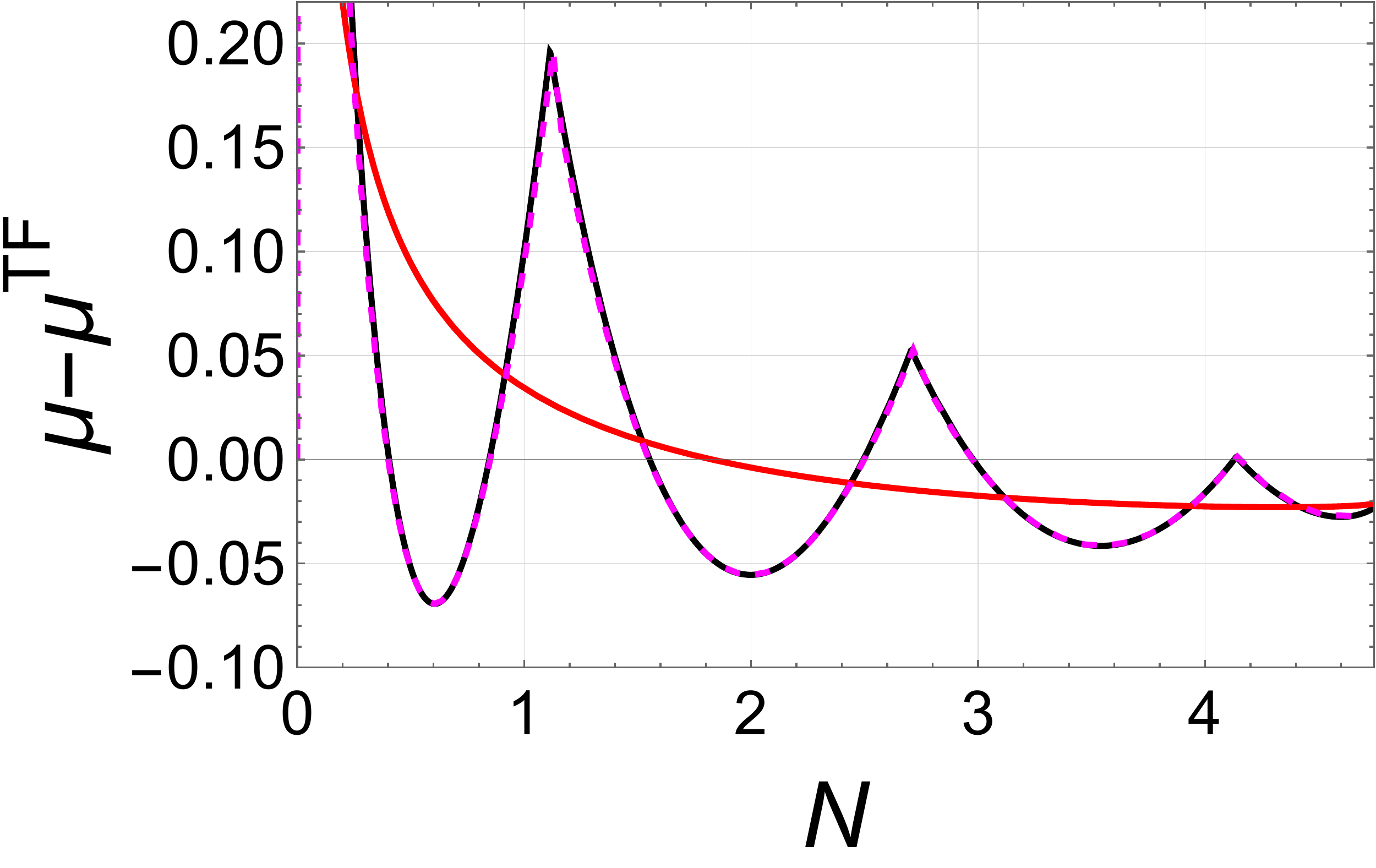}
\caption{Upper panel: The exact energy (black) and its AEA2 (red) and AEA4 (blue) approximations (with the TF energy subtracted), for the $M = 5$ PT slab in Table \ref{tab:PTWDMuDCon}.  The middle panel shows a zoomed in view of the upper panel.  The gray line marks the particle number we examined in Table \ref{tab:PTWDMuDCon}.  Lower panel: The exact chemical potential for a PT slab with $D = 10$ (black), $dE\AEAtwo/dN$ (red), and $\mu\AEAtwo(N)$ (magenta).  We have subtracted the TF value from all curves.}
\label{fig:ENL}
\end{figure}

In Table \ref{tab:IPAndMuTrunc} we calculate electron removal energies (RE) in two distinct ways.  To simulate electron removal in a molecule, we introduce $e_{RE}$: the energy per electron to remove $0.5$ electrons (each per unit area).  In fact $\mu$ is the true ionization potential of a metal, which $e_{RE}$ approaches in the LS limit.  The left-hand errors are those found from energy differences, in which AEA2 is surprisingly poor.  This is because AEA2 for $E(N)$ has no oscillations (the cusps in Fig. \ref{fig:TNL} are in $T$, which does).  On the other hand, $\mu\AEAtwo(N)$ does contain quantum oscillations, as shown in the lower panel, which plots the exact chemical potential and two approximations to it.  The right-hand side of the table shows much better results from $\mu\AEAtwo(N)$, illustrating the subtleties of derivatives of the oscillating terms.  The derivatives of oscillations are typically one order larger than those of the corresponding smooth terms. 

Finally, Fig. \ref{fig:ENL} shows the exact and two approximate energy curves, each with the TF curve subtracted, as a function of $N$.  The black curve (exact) is oscillating, as is the blue (AEA4) approximation.  But the red curve (AEA2) contains no oscillations, as the oscillations cancel out of the total energy curve to 2nd order [which means that $E\AEAtwo(N) = E^{\rm{AEA2'}}(N)$].  The middle panel in Fig. \ref{fig:ENL} zooms in on the energy curve in the region of the particle number corresponding to $\mu=D/2$, marked by the vertical line.  Clearly the oscillations play a large role in the energy change if you remove 1/2 an electron when $\mu = D/2$.  The red curve yields a very poor approximation to this energy difference.  The exact energy satisfies $E'(N) = \mu$, but $dE\AEAtwo/dN$ yields a poor approximation to the chemical potential.  Instead we derive $\mu\AEAtwo(N)$ by inverting $N\AEAtwo(\mu)$.  The blue AEA4 curve will clearly yield almost exact answers.

\subsection{Binding energies}
\begin{figure}[htb!]
\includegraphics[width=0.95\columnwidth]{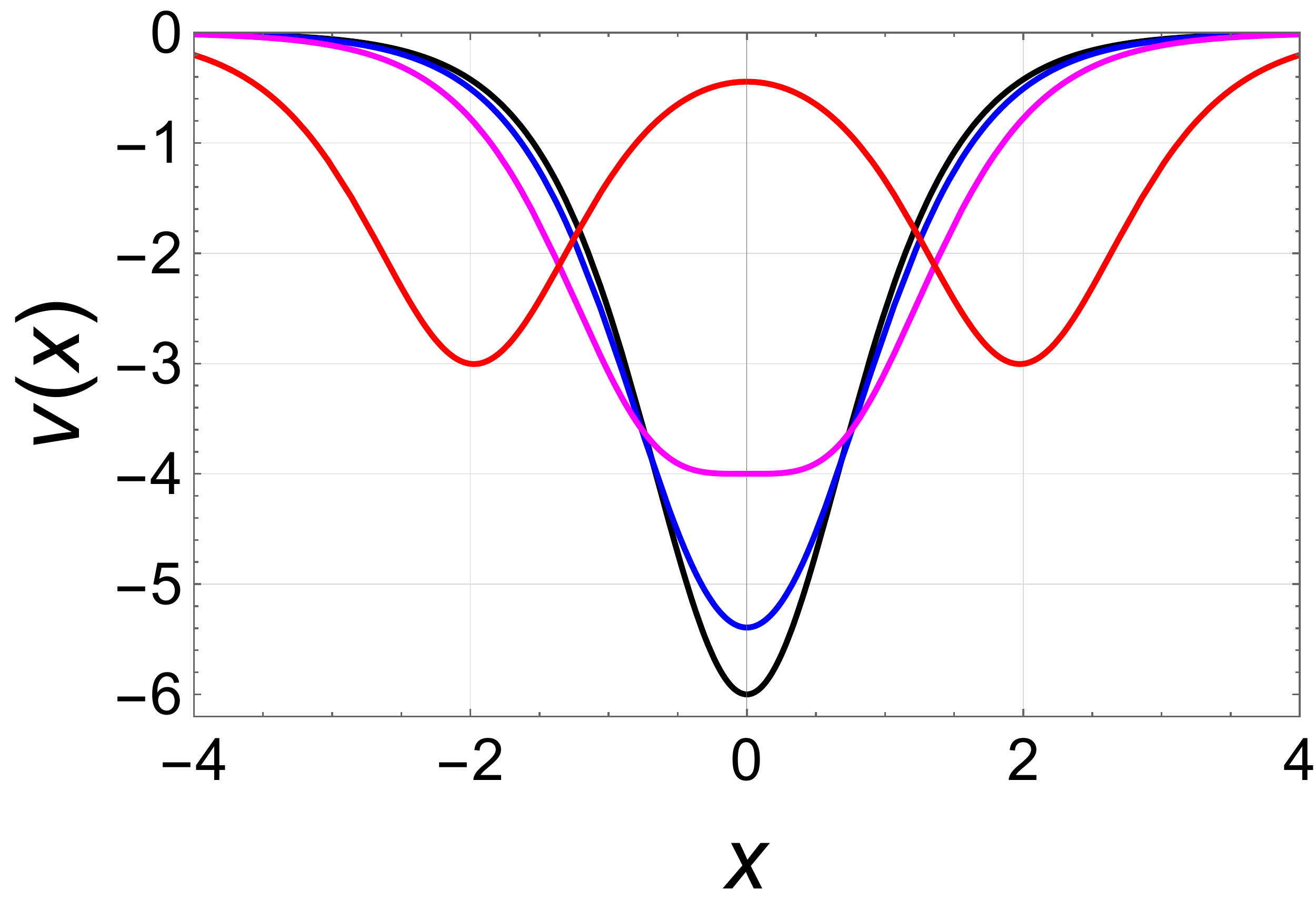}
\caption{Several PT dimer potentials made from two PT slabs (each with $D = 3$) at various separations: $R/R_c = 0$, 0.5, 1, 3 (black, blue, magenta, red), where $R_c= 2\, \asech \sqrt{2/3}$ and $v(\pm \infty) = 0$.}
\label{fig:BondBreak}
\end{figure}

Our last test is the most stringent.  We consider the sum of two identical PT slabs (the PT dimer), each of well depth $D = 3$, as a function of their separation, $R$, to mimic the binding energy curve of a molecule.  From 0 to $R_c = 2\, \asech \sqrt{2/3} = 1.31696$ the depth of each dimer is given by $\mathcal{D} = 6\,\sech^2(R/2)$ and there is a single minimum.  Beyond $R_c$, there is a double well and the formula for the total well depth becomes more complicated (see Fig. \ref{fig:BondBreak}).  We keep $N$ fixed throughout.  Table \ref{tab:TBind} shows the result.  For $R \leq R_c$, our results for energy differences are similar to, but less accurate than, those for total energies in Table \ref{tab:PTWDMuDCon}, with AEA4 still yielding chemical accuracy.  Figure \ref{fig:BondBreak} shows various potentials of the PT dimer slabs as a function of their separation, given in units of the critical separation ($R_c$) at which the second derivative of the midpoint potential vanishes.  Beyond this critical value, there are two wells, and our derivation of the semiclassical asymptotic expansion no longer applies.

\subsection{Delocalization error}

\begin{table*}
$\begin{array}{|c|c|c|c|c|c|r|r|c|c|c|}
\hline
\mc{2}{|c|}{} & \mc{9}{c|}{\text{Errors (mH)}}\\
\hline
\mc{2}{|c|}{} & \mc{5}{c|}{\text{Potential Functionals}} & \mc{4}{c|}{\text{Density Functionals}}\\
\hline
R/R_c & T - 2T_A & \text{TF} & \text{GEA2} & \text{AEA2'} & \text{AEA2} & \mc{1}{c|}{\text{AEA4}} & \mc{1}{c|}{\text{TF}} & \text{GEA2} & \text{MGE2} & \text{GEA4} \\
\hline
 0   & 0.792 & 27 & 28 & 32.2 & 6.1 & -0.177 & -23.8 & -11 & -7 & -12 \\
0.1  & 0.784 & 28 & 30 & 32.1 & 6.1 & -0.179 & -23.0 & -11 & -7 & -12 \\
0.2  & 0.760 & 31 & 34 & 31.7 & 6.2 & -0.186 & -20.6 & -10 & -7 & -12 \\
0.25 & 0.743 & 33 & 37 & 31.4 & 6.3 & -0.191 & -18.8 & -10 & -7 & -12 \\
0.3  & 0.722 & 35 & 41 & 31.0 & 6.2 & -0.197 & -16.8 &  -9 & -7 & -12 \\
0.4  & 0.672 & 38 & 48 & 30.0 & 6.1 & -0.211 & -11.8 &  -8 & -7 & -11 \\
0.5  & 0.613 & 41 & 55 & 28.5 & 5.7 & -0.219 &  -6.0 &  -6 & -7 & -10 \\
0.6  & 0.547 & 43 & 60 & 26.6 & 5.0 & -0.204 &   0.2 &  -5 & -6 & -9 \\
0.7  & 0.477 & 42 & 63 & 24.4 & 4.1 & -0.141 &   6.5 &  -2 & -5 & -8 \\
0.75 & 0.441 & 41 & 64 & 23.3 & 3.7 & -0.083 &   9.4 &  -1 & -4 & -7 \\
0.8  & 0.406 & 40 & 64 & 22.2 & 3.2 & -0.001 &  12.3 &   0 & -4 & -6 \\
0.9  & 0.337 & 35 & 62 & 20.1 & 2.5 &  0.251 &  17.3 &   2 & -3 & -4 \\
1    & 0.271 & 30 & 59 & 18.4 & 2.1 &  0.644 &  21.3 &   3 & -2 & -3 \\
1.1  & 0.211 & 24 & 54 & 17.2 & 2.2 &  1.191 &  24.0 &   4 & -2 & -2 \\
1.2  & 0.157 & 18 & 48 & 16.8 & 2.8 &  1.876 &  25.3 &   4 & -2 & -1 \\
1.25 & 0.133 & 15 & 45 & 16.8 & 3.4 &  2.253 &  25.5 &   4 & -2 & -1 \\
\hline
\end{array}$
\caption{The kinetic binding energies, $T - 2T_A$, for a series of PT dimer slabs with $3/\pi$ electrons per unit area.  $T$ is the exact kinetic energy and $T_A$ is the kinetic energy of an isolated PT slab ($D=3$).  All approximations are defined as in Table \ref{tab:PTWDMuDCon}.  We give more energy tables for these dimers in the appendix.} 
\label{tab:TBind}
\end{table*}

While AEA4 yields chemical accuracy for any $R \leq R_c$, it fails to do so beyond $R_c$.  This may be because a different, more complicated asymptotic expansion applies, one that accounts for the additional turning points below $\mu$ (there are still only two turning points at $\mu$).  The $R_c < R$ error in AEA2 remains below that of $R=0$, although it is also growing slightly.  This is characteristic of an asymptotic expansion and presumably, for $R$ sufficiently large, AEA2 would fail.  Eventually for $R$ very large, as the transition to four turning points at $\mu$ is approached, even TF should fail.  This is highly analogous to the situation for XC \cite{BCGP16}, but beyond the scope of the present work.

\section{Conclusions}
Our PFAs are the asymptotic expansion in $\hbar$ of the energy to a given order, derived directly from the Schr\"odinger equation.  Unlike any DFAs in use today, they are the correct generalization of the gradient expansion (as a PFA) to finite systems.  Our results bring the proof-of-principle from earlier 1D studies closer to realistic orbital-free electronic structure calculations.  The leading terms in the semiclassical AEA achieve much greater accuracy than modern orbital-free DFAs, and their quantum oscillations capture derivative discontinuities.  However, our results are limited to homogeneous slabs with only two turning points.

For such homogeneous slab calculations, our PFAs can be tested immediately.  With any existing KS code and XC choice, find the self-consistent solution, and then test our PFAs by feeding them the KS potential.  Pseudopotentials are smoother than full potentials, and so are likely to lead to higher accuracy for a given order.  To do this self-consistently, the XC potential requires the ground-state density.  There are highly accurate PFA's for the 1D density \cite{RLCE15,RB17,RB18}, which can be generalized to slabs and then tested for sufficient accuracy.  Otherwise one can extract a PFA density via $\rho(x)=\delta E/\delta v(x)$ for fixed $N$, a non-trivial derivation when oscillating terms are included.  Alternatively, the Lieb maximization principle, $T[\rho] = \sup\{E[v]-\int \rho\, v\}$, where the supremum is over all potentials \cite{CGB13,CLEB11}, converts a PFA into a DFA.  This non-trivial derivation for AEA2 has several caveats.  Since the AEA is not variational, such a maximization might not yield useful results (the bare AEA might need to be generalized).  Moreover, the resulting DFA might be less accurate than the original PFA for finite systems.

Another interesting area is surfaces.  As the interior region of a slab is broadened, quantum oscillations will become less relevant, and the AEA should reduce to the Airy-gas \cite{LMA14} approximation for $T_s$, and  AEA2 should give accurate surface kinetic energies.  As for inhomogeneity in the perpendicular directions,  there are many approximations one could try (e.g., application to surface averaged potentials, application in the direction of the local gradient of the potential, and simple GGA-type generalizations of AEA expressions) that could be constrained to recover our results in the uniform limit.  Another exciting area is to generalize the AEA to include complex turning points when bonds are stretched \cite{OB22}.  This has not yet been done, even in one-dimension. Real turning points determine the dominant form of the semiclassical expansion, and complex turning points yield subdominant corrections to this expansion.

Could these techniques be applied to the XC energy?  Yes, but with difficulty.  Consider exchange.  Our most accurate results come from fitting highly accurate atomic data for atoms up to $Z =$ 120, because we lack analytic results \cite{ARCB22}.  Asymptotic expansions must be found for the double sum in the exchange energy, and one also needs the semiclassical expansion for the KS density matrix with real turning points \cite{ECPG15}.  A computational nirvana might be reached \cite{BB20} if we could overcome that one difficulty.  

We thank the NSF (CHE-2154371) for funding. More rigorous derivations, exact results for PT slabs, and limiting cases, will appear in future work.

\bibliography{Bib}

\appendix
\section*{Appendix: Tables of energies}
\label{sec:Eslabs}
This appendix contains tables that supplement those in the main text.

\begin{table*}
$\begin{array}{|c|r|r|r|c|c|c|c|c|c|c|}
\hline
\mc{4}{|c|}{} & \mc{7}{c|}{\text{Errors (mH)}}\\
\hline
\mc{4}{|c|}{} & \mc{3}{c|}{\text{Potential Functionals}} & \mc{4}{c|}{\text{Density Functionals}}\\
\hline
M & \mc{1}{|c|}{D} & \mc{1}{c|}{N} & \mc{1}{c|}{E/N} & \text{TF} & \text{AEA2} & \text{AEA4} & \text{TF} & \text{GEA2} & \text{MGE2} & \text{GEA4}
\\
\hline
1  &  12.685 &   1.293 &   4.312 & -192 & 9.2 & 0.003781 & -156 & -41 & -8 & -2 \\
2  &  36.000 &   6.525 &  12.189 & -190 & 3.1 & 0.000452 & -159 & -35 &  1 & -6 \\
3  &  70.971 &  18.318 &  24.005 & -189 & 1.6 & 0.000115 & -162 & -31 &  7 & -7 \\
4  & 117.599 &  39.293 &  39.759 & -189 & 0.9 & 0.000042 & -164 & -28 & 11 & -6 \\
5  & 175.883 &  72.075 &  59.451 & -189 & 0.6 & 0.000019 & -165 & -26 & 14 & -6 \\
6  & 245.824 & 119.288 &  83.082 & -189 & 0.4 & 0.000010 & -166 & -25 & 16 & -6 \\
7  & 327.422 & 183.555 & 110.651 & -189 & 0.3 & 0.000005 & -167 & -24 & 18 & -6 \\
8  & 420.677 & 267.500 & 142.159 & -189 & 0.3 & 0.000003 & -168 & -23 & 19 & -6 \\
9  & 525.589 & 373.746 & 177.605 & -189 & 0.2 & 0.000002 & -168 & -22 & 21 & -5 \\
10 & 642.157 & 504.918 & 216.990 & -189 & 0.2 & 0.000001 & -169 & -21 & 22 & -5 \\
\hline
\end{array}$
\caption{Same as Table \ref{tab:PTWDMuDCon} of the main text, but for the total energy.}
\label{tab:PTWDMuDConE}
\end{table*}
Table \ref{tab:PTWDMuDConE} shows the total energies (not just the kinetic energies) of the calculations in Table \ref{tab:PTWDMuDCon} of the main text.  In this case, any functionals evaluated on the exact density include the exact potential energy by construction.  Just as in the self-consistent TF calculation, we expect errors on (some version of) self-consistent densities to be larger.

\begin{table*}
$\begin{array}{|c|c|c|c|r|r|r|c|c|c|c|}
\hline
\mc{2}{|c|}{} & \mc{9}{c|}{\text{Errors (mH)}}\\
\hline
\mc{2}{|c|}{} & \mc{5}{c|}{\text{Potential Functionals}} & \mc{4}{c|}{\text{Density Functionals}}\\
\hline
R/R_c & T & \text{TF} & \text{GEA2}& \mc{1}{c|}{\text{AEA2'}} & \mc{1}{c|}{\text{AEA2}} & \mc{1}{c|}{\text{AEA4}} & \text{TF} & \text{GEA2} & \text{MGE2} & \text{GEA4} \\
\hline
0   & 1.890 & 34 & -19 & 11.8 &  3.1 & -0.038 & -77 & -22 & -5.5 & -12 \\
0.1  & 1.882 & 35 & -17 & 11.7 &  3.1 & -0.041 & -76 & -21 & -5.5 & -12 \\
0.2  & 1.858 & 38 & -13 & 11.4 &  3.2 & -0.047 & -74 & -21 & -5.6 & -12 \\
0.25 & 1.841 & 40 & -10 & 11.1 &  3.2 & -0.052 & -72 & -21 & -5.6 & -12 \\
0.3  & 1.821 & 41 &  -6 & 10.7 &  3.2 & -0.058 & -70 & -20 & -5.6 & -11 \\
0.4  & 1.771 & 45 &   1 &  9.6 &  3.1 & -0.072 & -65 & -19 & -5.4 & -11 \\
0.5  & 1.711 & 48 &   8 &  8.1 &  2.6 & -0.080 & -59 & -17 & -5.0 & -10 \\
0.6  & 1.645 & 50 &  13 &  6.2 &  2.0 & -0.065 & -53 & -15 & -4.3 &  -9 \\
0.7  & 1.575 & 49 &  17 &  4.1 &  1.1 & -0.002 & -47 & -13 & -3.4 &  -7 \\
0.75 & 1.540 & 48 &  17 &  2.9 &  0.6 &  0.056 & -44 & -12 & -2.8 &  -6 \\
0.8  & 1.504 & 47 &  17 &  1.8 &  0.2 &  0.138 & -41 & -11 & -2.3 &  -6 \\
0.9  & 1.435 & 42 &  16 & -0.3 & -0.6 &  0.390 & -36 &  -9 & -1.3 &  -4 \\
1    & 1.369 & 37 &  12 & -2.0 & -1.0 &  0.783 & -32 &  -8 & -0.6 &  -3 \\
1.1  & 1.309 & 31 &   7 & -3.2 & -0.9 &  1.330 & -29 &  -7 & -0.1 &  -2 \\
1.2  & 1.255 & 25 &   1 & -3.6 & -0.2 &  2.014 & -28 &  -6 & -0.2 &  -1 \\
1.25 & 1.231 & 22 &  -2 & -3.6 &  0.3 &  2.392 & -28 &  -6 & -0.3 &  -1 \\
\hline
\end{array}$
\caption{Same as Table \ref{tab:TBind} of the main text, but showing the total kinetic energies, not the kinetic binding energies.}
\label{tab:2PTT}
\end{table*}

\begin{table*}
$\begin{array}{|c|r|c|r|r|r|c|c|c|}
\hline
\mc{2}{|c|}{} & \mc{7}{c|}{\text{Errors (mH)}}\\
\hline
\mc{2}{|c|}{} & \mc{3}{c|}{\text{Potential Functionals}} & \mc{4}{c|}{\text{Density Functionals}}\\
\hline
R/R_c & \mc{1}{c|}{E - 2E_A} & \text{TF} & \text{AEA2} & \mc{1}{c|}{\text{AEA4}} & \mc{1}{c|}{\text{TF}} & \text{GEA2} & \text{MGE2} & \text{GEA4} \\
\hline
0    &  1.174 & -39 & -8.1 &  0.14 & -23.8 & -11 & -7.1 & -12 \\
0.1  &  1.162 & -39 & -8.2 &  0.14 & -23.0 & -11 & -7.1 & -12 \\
0.2  &  1.127 & -36 & -8.5 &  0.14 & -20.6 & -10 & -7.2 & -12 \\
0.25 &  1.100 & -34 & -8.6 &  0.14 & -18.8 & -10 & -7.2 & -12 \\
0.3  &  1.068 & -32 & -8.7 &  0.14 & -16.8 &  -9 & -7.2 & -12 \\
0.4  &  0.988 & -27 & -8.7 &  0.13 & -11.8 &  -8 & -7.0 & -11 \\
0.5  &  0.887 & -21 & -8.3 &  0.10 &  -6.0 &  -6 & -6.6 & -10 \\
0.6  &  0.767 & -13 & -7.4 &  0.04 &   0.2 &  -5 & -5.9 &  -9 \\
0.7  &  0.630 &  -6 & -6.0 & -0.04 &   6.5 &  -2 & -5.0 &  -8 \\
0.75 &  0.556 &  -2 & -5.1 & -0.09 &   9.4 &  -1 & -4.5 &  -7 \\
0.8  &  0.479 &   2 & -4.2 & -0.14 &  12.3 &   0 & -3.9 &  -6 \\
0.9  &  0.315 &   9 & -2.3 & -0.23 &  17.3 &   2 & -3.0 &  -4 \\
1    &  0.143 &  15 & -0.5 & -0.28 &  21.3 &   3 & -2.2 &  -3 \\
1.1  & -0.036 &  19 &  0.7 & -0.29 &  24.0 &   4 & -1.8 &  -2 \\
1.2  & -0.217 &  22 &  1.3 & -0.21 &  25.3 &   4 & -1.8 &  -1 \\
1.25 & -0.309 &  22 &  1.2 & -0.12 &  25.5 &   4 & -1.9 &  -1 \\
\hline
\end{array}$
\caption{Same as Table \ref{tab:TBind} of the main text, but with the total binding energy.}
\label{tab:BindE}
\end{table*}

\begin{table*}
$\begin{array}{|c|c|c|r|r|c|c|c|c|}
\hline
\mc{2}{|c|}{} & \mc{7}{c|}{\text{Errors (mH)}}\\
\hline
\mc{2}{|c|}{} & \mc{3}{c|}{\text{Potential Functionals}} & \mc{4}{c|}{\text{Density Functionals}}\\
\hline
R/R_c & E & \text{TF} & \mc{1}{c|}{\text{AEA2}} & \mc{1}{c|}{\text{AEA4}} & \text{TF} & \text{GEA2} & \text{MGE2} & \text{GEA4} \\
\hline
0    & 2.845 & -99 & -7.2 &  0.15 & -77 & -22 & -5.5 & -12 \\
0.1  & 2.833 & -98 & -7.3 &  0.16 & -76 & -21 & -5.5 & -12 \\
0.2  & 2.798 & -95 & -7.6 &  0.16 & -74 & -21 & -5.6 & -12 \\
0.25 & 2.772 & -94 & -7.7 &  0.16 & -72 & -21 & -5.6 & -12 \\
0.3  & 2.740 & -92 & -7.8 &  0.16 & -70 & -20 & -5.6 & -11 \\
0.4  & 2.659 & -86 & -7.8 &  0.15 & -65 & -19 & -5.4 & -11 \\
0.5  & 2.558 & -80 & -7.4 &  0.12 & -59 & -17 & -5.0 & -10 \\
0.6  & 2.438 & -73 & -6.5 &  0.06 & -53 & -15 & -4.3 &  -9 \\
0.7  & 2.301 & -65 & -5.1 & -0.03 & -47 & -13 & -3.4 &  -7 \\
0.75 & 2.227 & -61 & -4.3 & -0.07 & -44 & -12 & -2.8 &  -6 \\
0.8  & 2.150 & -57 & -3.3 & -0.12 & -41 & -11 & -2.3 &  -6 \\
0.9  & 1.986 & -51 & -1.4 & -0.21 & -36 &  -9 & -1.3 &  -4 \\
1    & 1.814 & -45 &  0.4 & -0.26 & -32 &  -8 & -0.6 &  -3 \\
1.1  & 1.636 & -40 &  1.6 & -0.27 & -29 &  -7 & -0.1 &  -2 \\
1.2  & 1.454 & -38 &  2.1 & -0.19 & -28 &  -6 & -0.2 &  -1 \\
1.25 & 1.362 & -37 &  2.1 & -0.10 & -28 &  -6 & -0.3 &  -1 \\
\hline
\end{array}$
\caption{Same as Table \ref{tab:TBind} of the main text, but with the total energy.  We fix the center of each well at 0: $v(0) = 0$.}
\label{tab:2PTE}
\end{table*}
Tables \ref{tab:2PTT}-\ref{tab:2PTE} supplement Table \ref{tab:TBind} of the main text, showing total kinetic energies of the PT dimer slabs, not just binding energies, so that approximations cannot benefit from cancellation of errors between the PT dimer slabs and the separated `atomic' slabs.  We also give the corresponding total energy and binding energy.

\end{document}